\documentclass[12pt]{article}

\usepackage{newtxtext,newtxmath}

\usepackage{graphicx}
\usepackage{float}
\usepackage[letterpaper,margin=1in]{geometry}

\renewenvironment{abstract}
	{\quotation}
	{\endquotation}

\date{}

\newcommand{\YS}[1]{{\color{black} #1}}

\makeatletter
\renewcommand{\fnum@figure}{\textbf{Figure \thefigure}}
\renewcommand{\fnum@table}{\textbf{Table \thetable}}
\makeatother

\usepackage{scicite}

\usepackage{url}

\def\scititle{
	Shape Anisotropy Governs Organization of Active Rods: Swarming, Turbulence, Flocking, and Jamming
}
\title{\bfseries \boldmath \scititle}

\author
{Yogesh Shelke,$^{1}$ Anpuj Nair S,$^{1}$ and Hanumantha Rao Vutukuri$^{1}$$^\ast$\and
\small{$^{1}$Active Soft Matter and Bio-inspired Materials Lab, Faculty of Science and Technology,}\and
\small{MESA+ Institute, University of Twente, The Netherlands}\and
\small{PO Box 217, 7500 AE Enschede, The Netherlands}\and
	\small$^\ast$Corresponding author. Email: h.r.vutukuri@utwente.nl
}

\begin{document} 


\maketitle 
\begin{abstract}\bfseries \boldmath

Shape anisotropy of individual building blocks plays a crucial role in creating exotic structures and controlling phase behavior in equilibrium systems. We present a combined experimental and simulation study using light-driven self-propelled rods to investigate when and how shape-induced alignment, steric, and hydrodynamic interactions govern self-organization. By varying rod aspect ratio and area fraction, the system evolves from active Brownian motion to swarming, active turbulence, flocking, large clusters, and jamming. A state diagram summarizes emergent behaviors, and spatiotemporal analyses reveal distinct giant-number fluctuations across states. This minimal model offers insight into the self-organization of biological rod-like microswimmers, enabling the decoupling of physical from biological mechanisms. Our results provide design rules for programmable synthetic active materials and highlight parallels with bacterial swarms and other biological assemblies.

\end{abstract}

\noindent Micron-sized self-propelled particles are of significant interest in active soft matter due to their ability to mimic the collective behaviors of living systems and serve as minimal models for studying intrinsically out-of-equilibrium dynamics \cite{bechingerActiveParticlesComplex2016,Gompper_2025}. At the microscale, where inertial effects are negligible, collective motion predominantly emerges from propulsion coupled with steric and hydrodynamic interactions, both of which are strongly influenced by particle shape. Synthetic spherical active Brownian particles have been studied in both experiments and simulations, demonstrating clustering \cite{yan2016reconfiguring,theurkauff2012dynamic,vutukuri2020active,palacciLivingCrystalsLightActivated2013} and motility-induced phase separation\cite{catesMotilityInducedPhaseSeparation2015}. However, unlike spherical swimmers, elongated self-propelled particles inherently exhibit anisotropic alignment due to their shape and experience orientation-dependent steric and hydrodynamic interactions, resulting in fundamentally different collective dynamics \cite{weitzSelfpropelledRodsExhibit2015,lushiFluidFlows2014}, as reported in several bacterial systems such as {\it Escherichia coli} \cite{colinChemotacticBehaviourEscherichia2019}, {\it Myxococcus xanthus} \cite{peruaniCollectiveMotionNonequilibrium2012}, and {\it Bacillus subtilis} \cite{zhangCollectiveMotionDensity2010}.

Several theoretical and simulation studies have predicted that elongated active particles, such as self-propelled rods, can exhibit a rich spectrum of emergent behaviors including swarming\cite{zantopEmergentCollectiveDynamics2022, peruaniNonequilibriumClusteringSelfpropelled2006}, polar flocks\cite{grossmann2020particle,broker2024collective}, nematic ordering\cite{ginelliLargeScaleCollectiveProperties2010, baskaran2008hydrodynamics}, active turbulence\cite{zantopEmergentCollectiveDynamics2022,alertActiveTurbulence2022,wensinkMesoscaleTurbulenceLiving2012}, and jamming\cite{wensinkMesoscaleTurbulenceLiving2012}. However, most experimental studies to date have been limited to dry granular rod systems \cite{narayanLongLivedGiantNumber2007, kudrolli2008swarming}, where the absence of hydrodynamic interactions and thermal fluctuations restricts their ability to mimic and elucidate the rich dynamics observed in biological systems (e.g., {\it E. coli} \cite{colinChemotacticBehaviourEscherichia2019}, {\it M. xanthus} \cite{peruaniCollectiveMotionNonequilibrium2012}, {\it B. subtilis }\cite{zhangCollectiveMotionDensity2010}). In contrast, micron-sized self-propelled rods represent a compelling minimal model due to their length scales and hydrodynamic flow fields comparable to those of biological counterparts \cite{Gompper_2025, palacciLivingCrystalsLightActivated2013, vutukuri2020light}. Moreover, they can effectively mimic a wide range of biological systems, including microtubules, sperm, {\it E. coli}, and biopolymers \cite{barSelfPropelledRodsInsights2020, yangSwarmBehaviorSelfpropelled2010}, and can be used to unravel the underlying principles that govern their complex collective behaviors. 
Colloidal side-propelling rods have been shown to form small clusters \cite{vutukuriDynamicSelforganizationSidepropelling2016}, whereas studies on metallic rods have largely focused on single particle behavior \cite{paxtonCatalyticNanomotorsAutonomous2004, mu2022light}. However, most existing synthetic rod systems do not fully capture key features of biological rod-like swimmers, particularly their micron-scale dimensions and propulsion along the long axis. Developing minimal model systems is therefore needed, not only to better understand self-organization in biological microswimmers (e.g., {\it E. coli}, {\it M. xanthus}, and {\it B. subtilis}), but also to decouple physical mechanisms (e.g., hydrodynamics and steric interactions) from biological ones (e.g., quorum sensing).

We present a combined experimental and active Brownian dynamics simulation study to unravel the collective dynamics of polar self-propelled colloidal rods. We systematically investigate how shape-induced alignment, arising from steric, hydrodynamic and phoretic interactions, governs their collective behavior. By independently varying the rod aspect ratio and area fraction, we identify a sequence of phases including swarming, active turbulence, flocking, phase separation, large clusters, and jamming. We consolidate these results into a state diagram that maps these phases as a function of rod aspect ratio and concentration of active rods.
\\

{\noindent\bf{Collective motion of polar active rods}\label{subsec2}}\\
We synthesized light-driven colloidal rods composed of a titania (\(\mathrm{TiO}_{2}\)) head and a silica (\(\mathrm{SiO}_{2}\)) tail (see Fig. \ref{fig:1}C inset and fig. S1), following the procedure described in \cite{mu2022light, kuijk2011synthesis} (see methods in SI). These rods serve as a tunable model system in which both aspect ratio and concentration can be controlled. As the rods are heavier than the dispersing medium, they sediment to the bottom of the observation chamber (see SI section 2.1), where their Brownian motion is observed in a quasi-2D plane under bright-field illumination. When exposed to green light ($\lambda \approx$ 532\text{–}552\,\text{nm}) with an intensity of $\sim$34 mW/cm\textsuperscript{2}, a light-induced redox reaction is triggered between fuel (hydroquinone (HQ) and benzoquinone (BQ)) at the (\(\mathrm{TiO}_{2}\)) head, generating local chemical gradients that drive propulsion along the rod’s long axis, in the direction of the head\cite{mu2022light}.

We begin by presenting experimental results for rods with aspect ratio $\alpha = l/d = 7.5$ (where $l = 5.3$\,$\mu$m and $d = 0.7$\,$\mu$m). By varying the area fraction $\phi$, we identified a sequence of distinct collective states (see Fig.~\ref{fig:1}C–F). At low area fractions (\(\phi \leq 0.06\)), rods propel with a strength of $3.7 \pm 0.5$ $\mu$m/$\text{s}$, showing active Brownian motion with $Pe=40$ (see SI section 2.1), and independently explore space with random orientations, as shown in the overlaid trajectory in Fig. \ref{fig:1}C and movie S1. As the area fraction increased to $\phi = 0.11$, rods began to encounter each other more frequently, and steric, hydrodynamic and phoretic interactions induced alignment, self-organizing the rods into small clusters of two to five rods that moved together. We visualized the fluid flows around immobilized rods using silica tracer particles (zeta potential of $\zeta=-62 ~mV$) in a fuel solution (Fig. \ref{fig:1}A) to understand how hydrodynamic interactions contribute to cluster formation. The streamlines and their magnitude, obtained from particle image velocimetry (PIV)\cite{thielicke2021particle}, reveal that fluid is consistently pushed outward from the front of the head and drawn inward to the back of the head, and flowing along the tail. This flow pattern is driven by the chemical gradients originating from the photocatalytic reaction at the rod's head, indicating a pusher-type propulsion mechanism, similar to those observed in biological systems \cite{
lushiFluidFlows2014,wensinkMesoscaleTurbulenceLiving2012,marchetti2013hydrodynamics}.

Furthermore, our quantitative analysis reveals that the velocity of tracer particles $v_r$ decays with distance $r$ as $v_r(r) = A/r^3$, with a dipole strength of  \( A = 36.9~\mu\mathrm{m}^4\,\mathrm{s}^{-1} \) (see Fig.~\ref{fig:1}B, fig. S7, and SI section 2.3 for more details). This scaling is consistent with theoretical predictions when phoretic interactions and osmotic flows are taken into account\cite{andersonColloidTransportInterfacial1989, spagnolieHydrodynamicsSelfpropulsionBoundary2012, drescherFluidDynamicsNoise2011}.
To further decouple phoretic-driven effects from osmotic contributions, we coated tracer particles with a pNIPAM brush layer of thickness $\sim45$ nm, following the method from \cite{carrasco2025characterization} (see SI sections 1.5 and 2.4). These polymer brushes suppress phoretic motion by screening solute–surface interactions in the interfacial layer, thereby decreasing the effective phoretic mobility. Under these conditions, tracer motion was dominated by osmotic outflow from the rod heads, with no inward tracer motion observed behind the heads of the rods (Fig.~\ref{fig:1}B, fig. S10B, and movie S2).

\begin{figure}[htbp!]
	\centering
	\includegraphics [width=0.8\textwidth]{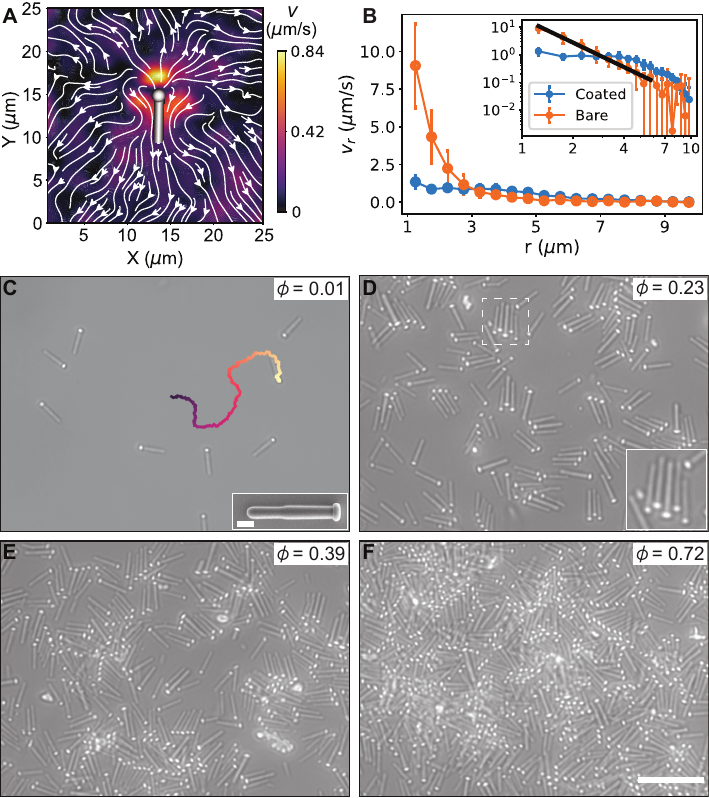} 
	\caption{\textbf{Collective motion of active colloidal rods with aspect ratio $\alpha=7.5$}. ({\bf A})
    Flow streamlines and magnitude indicate that fluid is strongly pushed away at the rod’s head, while streamlines curve inward behind the head and dissipate along the tail, confirming that the rods exhibit pusher-type behavior. ({\bf B}) Decay of radial component of tracer velocities, $v_r$ vs distance $r$ in front of rod-head for bare silica ($\sigma = $ 0.95~$\mu$m) and pNIPAM coated ($\sigma = $ 1.04~$\mu$m) silica tracers. Inset: Log-log plot of $v_r$ vs $r$ showing a scaling close to $1/r^3$ in the case of bare tracers. The black line represents a fit with exponent -2.9.  Error bars represent standard deviation across four rods. ({\bf C}- {\bf F}) Representative combined bright-field and fluorescence microscope images revealing different collective states with increasing area fraction ($\phi$): ({\bf C}) \(\phi = 0.01\), isotropic state. The inset shows the scanning electron microscope image of a rod with a  $\mathrm{TiO_2}$ head and $\mathrm{SiO_2}$ tail. ({\bf D}) \(\phi = 0.23\), swarming state. The inset shows the laterally staggered or zigzag arrangement of rods within the cluster. ({\bf E}) \(\phi = 0.39\), turbulence state. ({\bf F}) \(\phi = 0.72\), large cluster state. Scale bar represents 10 $\mu$m, and the inset is 1 $\mu$m.} 
   
	\label{fig:1}
\end{figure}

When two rods approach each other in a head-to-head configuration with opposite propulsion directions, they experience hydrodynamic repulsion (see SI section 2.5), typically resulting in transient interactions or separation (see fig. S12). In contrast, when rods propel in the same direction and approach each other at an acute angle within a distance of 4-5$\, \mu\mathrm{m}$, their lateral flow fields overlap and induce hydrodynamic attraction, leading to the formation of pairs (see Fig.~\ref{fig:1}A and fig. S11).
Extending beyond pairwise interactions, shape-induced effects influence the emergence and evolution of clusters comprising many rods. For instance, overlaid trajectories reveal the dynamic nature of these clusters, which transition from disordered to coordinated motion and then revert back to uncoordinated motion (see movie S3). Within these clusters, individual rods synchronize their orientations with their neighbors, as quantified by the local polar order parameter (see fig. S13 and movie S3). 
Although all rods remain aligned and propel in the same direction, the distance between neighboring rods fluctuates in an uncorrelated manner (see fig.~S14), indicating 
the absence of geometric locking. Within the swarm, the rods exhibit a staggered arrangement, as shown in the inset of Fig.~\ref{fig:1}D. This configuration arises from effective fluid-mediated hydrodynamic repulsion, which suppresses side-by-side head alignment, combined with a weak phoretic attraction at the back of the head of the rod. Using rods with identical head and tail sizes, we demonstrate that swarming is not a consequence of the slightly larger head, but rather a general feature of our polar active rods (see fig.~S15).

\begin{figure}[htbp!]
	\centering
	\includegraphics [width=0.95\textwidth]{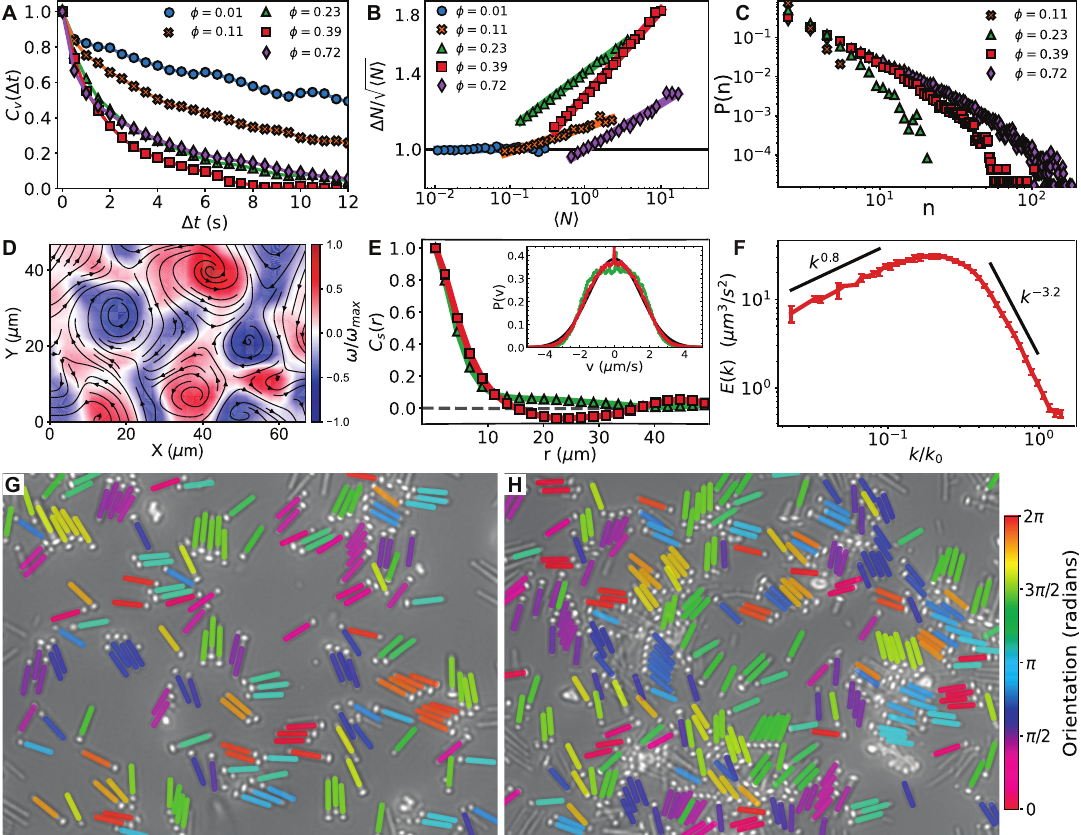} 
	\caption{\textbf{Spatiotemporal analysis of active rods with aspect ratio $\alpha = 7.5$ in different states} ({\bf A}) Velocity autocorrelation function (VACF) for rods at various area fractions $\phi$. A comparison across all states shows that the VACF decays most rapidly in the turbulent state ($\phi = 0.39$), where rods frequently change direction due to chaotic swirling flows.
({\bf B}) Normalized number fluctuations plotted against the average number of rods for different collective states, revealing that the turbulent state exhibits the largest scaling. The horizontal solid black line at $\Delta N/\sqrt{\langle N \rangle} = 1$ represents the baseline expected for an isotropic (equilibrium-like) system.
({\bf C}) Cluster size distributions $P(n)$ for rods at different area fractions $\phi$. ({\bf D}) Normalized vorticity field showing clockwise (blue) and counterclockwise (red) vortices in the turbulent state, corresponding to Fig.~\ref{fig:1}E.
({\bf E}) Comparison of spatial velocity correlations and normalized distributions of Cartesian velocity components (inset) for swarming (green) and turbulent (red) states. The distributions are normalized by subtracting their respective means and dividing by the standard deviation. The black line in the inset represents a Gaussian distribution. ({\bf F}) Energy spectra $E(k)$ in the turbulent state at $\phi\approx0.39$. $k_0=2\pi /L$ is the wave number associated with the rod length. Error bars represent standard deviation across four experiments. ({\bf G}– {\bf H}) Bright-field microscope images of active rods with overlaid color-coded orientation angles, highlighting local polar order in ({\bf G}) the swarming state at $\phi = 0.23$, and ({\bf H}) the turbulent state at $\phi = 0.39$.}
    
	\label{fig:2}
\end{figure}

A further increase in area fraction ($\phi = 0.23$) leads to the formation of cohesive swarms, where clusters of rods locally align their orientations and move collectively (Fig. \ref{fig:1}D, Fig. \ref{fig:2}G, and movie S4). This transition is reflected in the broadening of the cluster size distribution, indicating the emergence of intermediate-sized clusters as rod-rod interactions become more frequent (see Fig.~\ref{fig:2}C). At \(\phi = 0.39\), the system transitioned from a swarming phase to a chaotic state that resembles an active turbulent phase, where clusters of rods with the same orientation emerge and frequently collide, leading to rotating chaotic swirling motion in which clusters rapidly assemble and disintegrate (Fig.~\ref{fig:1}E, Fig.~\ref{fig:2}D and H, and see movie S5). This dynamic state is driven by hydrodynamic torques between rods, which destabilize local alignment and generate rotational stresses, thereby resulting in vortex-like structures.  The cluster size distribution broadened significantly with a wider range of cluster sizes (Fig. \ref{fig:2}C). As we further increased the area fraction of the rods to \(\phi = 0.72\), larger and stable slow moving clusters formed, as evidenced by the shift of the cluster size distribution toward higher values (Fig.~\ref{fig:2}C). In this state, steric hindrance restricted individual rod movement, leading to a reduction in velocity (fig. S5), yet cooperative motion was still observed within these dense clusters (Fig.~\ref{fig:1}F and movie S6). Finally, at \(\phi \approx 0.9\), due to the effects of crowding, the velocity decreased further, and the system reached a fully jammed state, where the densely packed rods became immobilized (fig. S2 and movie S7).
\\

{\noindent\bf{Spatiotemporal dynamics}\label{subsec2}}\\
To better understand the temporal and spatial dynamics of active rods across different phases, we analyzed the velocity autocorrelation function (VACF) and number fluctuations. Figure~\ref{fig:2}A compares the VACF curves for isotropic, swarming, turbulent, and large-cluster states. In the isotropic phase ($\phi = 0.01$), the VACF decays slowly, reflecting minimal interactions and persistent motion. In the swarming regime ($\phi = 0.11$), the decay becomes more rapid due to the onset of steric and hydrodynamic interactions that cause local reorientations, despite coherent group motion. At a higher area fraction ($\phi = 0.23$), stronger swarming leads to an even faster decay. The turbulent phase ($\phi = 0.39$) exhibits the fastest decay, driven by chaotic local flows and strong hydrodynamic torques that frequently reorient particle velocities. In contrast, the large-cluster phase ($\phi = 0.72$) shows a steep initial drop due to steric confinement, followed by a slower decay at longer times, indicating the presence of dense, stable clusters.

To probe deviations from equilibrium behavior in polar active rods across different states, we measured how normalized number fluctuations (\(\Delta N / \sqrt{\langle N \rangle}\)) scale with the mean number of particles (\(\langle N \rangle\)) in subsystems of varying sizes (2–10 $\mu$m). This scaling relation is expressed as:
\( \Delta N / \sqrt{\langle N \rangle} \propto \langle N \rangle^\beta\)
where the exponent $\beta$ quantifies the degree of deviation from equilibrium statistics. In equilibrium systems, fluctuations follow the central limit theorem, yielding $\beta = 0$. Deviations from this scaling indicate non-equilibrium behavior: $\beta > 0$ corresponds to Giant Number Fluctuations (GNF) arising from correlated motion or clustering, while $\beta < 0$ reflects suppressed fluctuations due to crowding effects. Figure~\ref{fig:2}B illustrates the distinct scaling behaviors observed experimentally in different collective states. In the dilute limit (\(\phi = 0.01\)), the system follows equilibrium scaling with \(\beta \approx 0\). In the swarming state ($\phi = 0.11$)
small-scale clustering causes slight deviations from equilibrium, with an exponent of \(\beta = 0.05\). As the area fraction increases further (\(\phi = 0.23\)), pronounced number fluctuations emerge due to enhanced swarming dynamics, characterized by $\beta = 0.11$. In the turbulent state (\(\phi = 0.39\)), large-scale density variations and chaotic dynamics significantly amplify fluctuations, resulting in GNFs with \(\beta = 0.17\). At \(\phi = 0.72\), where large stable clusters dominate, fluctuations are partially suppressed relative to the turbulent regime, with \(\beta = 0.09\). This scaling behavior systematically evolves across states, reflecting changes in collective dynamics and spatial organization \cite{marchetti2013hydrodynamics}. Similar GNFs are known to arise in a variety of driven systems, including granular matter, bacterial colonies, and self-propelled spherical particles \cite{narayanLongLivedGiantNumber2007,zhangCollectiveMotionDensity2010,peruaniCollectiveMotionNonequilibrium2012,liuDensityFluctuationsEnergy2021}.

To gain further insight into the turbulent state of active rods with $
\alpha = 7.5$ at $\phi = 0.39$, we analyzed transient flow dynamics using PIV. The vorticity fields $\omega = \nabla \times \mathbf{v}$ and streamline plots (Fig.~\ref{fig:2}D) revealed the formation of counter-rotating vortices. Time-resolved vorticity analysis shows that rod clusters converge at $t = 0$ s, form vortices at $t = 1$ s, reach a fully developed state by $t = 2$ s, dissolve at $t = 4$ s, and then reform, indicating a highly dynamic process with a typical vortex lifetime of approximately 2–3 seconds (see fig. S18 and Movie S5). Next, we analyzed spatial velocity correlations, the distribution of velocity components (inset), and the energy spectra in Fig.~\ref{fig:2}. The energy spectra in Fig.~\ref{fig:2}F exhibit two distinct scaling regimes with exponents $0.8$ and $-3.2$, reflecting a characteristic energy distribution across length scales that is consistent with active turbulence reported in previous studies\cite{alertActiveTurbulence2022,wensinkMesoscaleTurbulenceLiving2012}. In the turbulent state (red), the spatial correlation function exhibits a pronounced negative correlation, with a maximum anti-correlation length of approximately 24 $\mu$m, which matches the typical vortex size observed in Fig.~\ref{fig:2}E and the peak in the energy spectra in Fig.~\ref{fig:2}F. The velocity distribution follows a Gaussian profile (inset), and all these features are hallmarks of active turbulence \cite{alertActiveTurbulence2022, qiEmergenceActiveTurbulence2022}. These observations underscore that our minimal synthetic model captures key features of bacterial turbulence, as seen in {\it E. coli} and {\it B. subtilis} \cite{pengImagingEmergenceBacterial2021, wensinkMesoscaleTurbulenceLiving2012}.
\\

{\noindent\bf{State diagram of polar active rods}\label{subsec2}}\\
To investigate how rod shape anisotropy influences state behavior, we synthesized active rods with four different aspect ratios, $\alpha = 4.5$, $7.5$, $11.6$, and $15.1$ (see fig. S1). The experimentally observed states were characterized using different descriptors, including cluster size distribution, velocity autocorrelation function (VACF), spatial velocity correlation function, number fluctuations, and energy spectra (see Fig. \ref{fig:2}). Based on these analyses, we constructed the state diagram shown in Fig. \ref{fig:4}, which reveals six distinct dynamical states: swarming, active turbulence, flocking, phase separation, large clusters, and jamming, as a function of aspect ratio $\alpha$ and area fraction $\phi$. 

\begin{figure}[htbp!]
	\centering
	\includegraphics [width=0.95\textwidth]{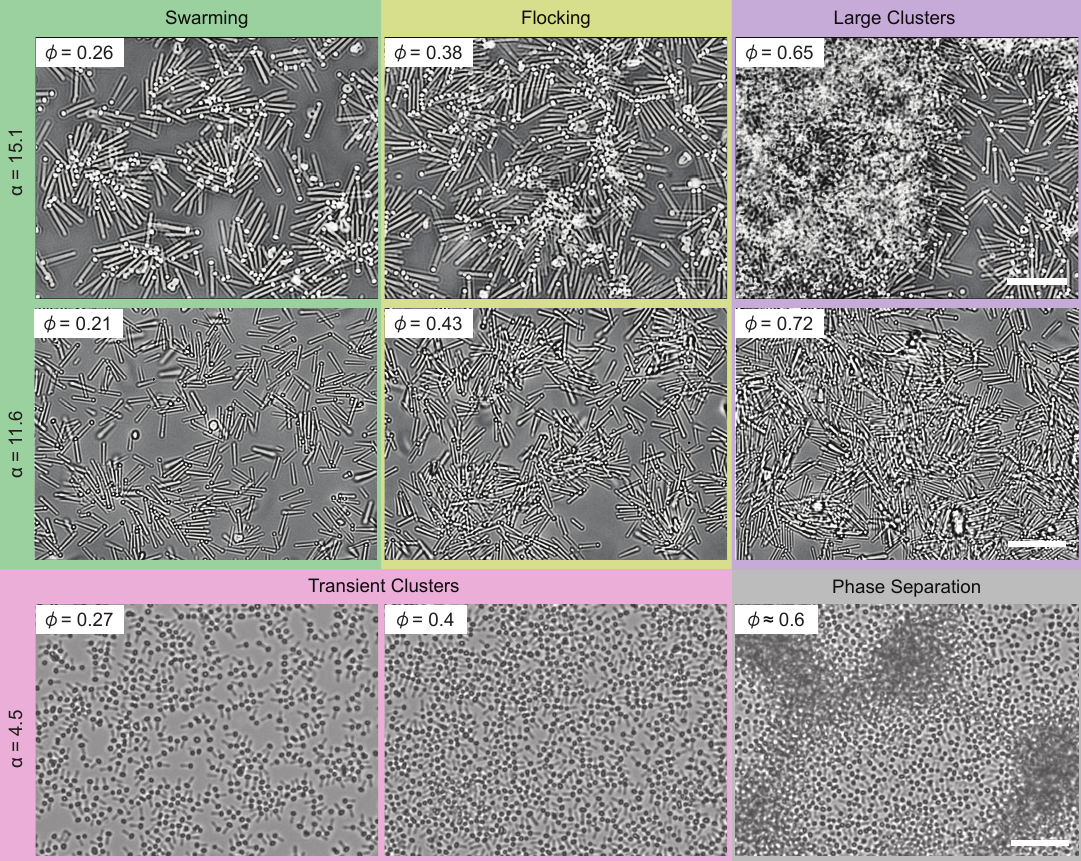} 
	\caption{\textbf{Combined bright-field and fluorescence microscope images showing emergent collective states for three different aspect ratios $\alpha$ across a range of area fractions.}  
    Short rods with $\alpha = 4.5$ give rise to a transient cluster state at intermediate area fractions due to their high rotational diffusion. At higher area fractions, their low anisotropy leads to phase separation into dilute and dense regions. In contrast, the longer rods ($\alpha = 11.6$ and $\alpha = 15.1$) exhibit distinct collective behaviors compared to $\alpha = 4.5$, namely swarming, flocking, and large-cluster states with area fraction. Images of rods with $\alpha = 11.6$ and $\alpha = 15.1$ were preprocessed with a band-pass filter to enhance contrast and visual clarity. Scale bar is 20 $\mu$m.}
	\label{fig:3}
\end{figure}

\begin{figure}[htbp!]
	\centering
	\includegraphics [width=0.99\textwidth]{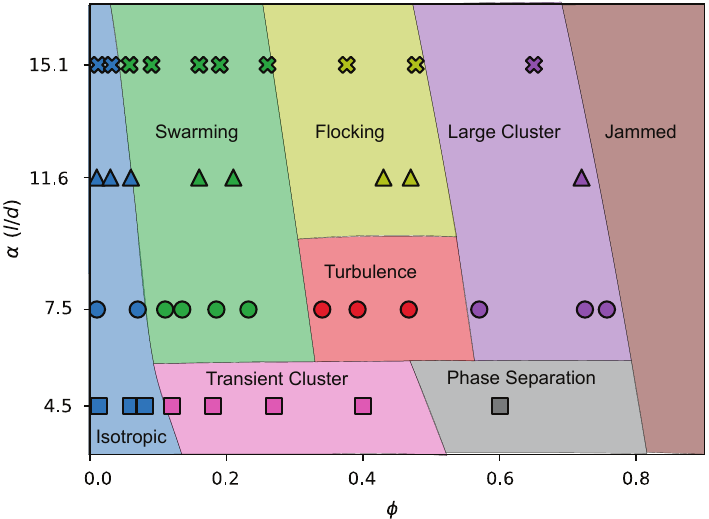}
	\caption{\textbf{Experimental state diagram of active rods.} Symbols denote different aspect ratios: squares for $\alpha = 4.5$, circles for $\alpha = 7.5$, triangles for $\alpha = 11.6$ and crosses for $\alpha = 15.1$. 
    Color-coded background regions serve as visual guides highlighting different collective states:
    Isotropic (Blue): randomly oriented rods with no collective motion; Swarming (Green): aligned rods exhibiting collective motion for $\alpha= 7.5$, $\alpha= 11.6$, and $\alpha = 15.1$.; Transient clusters (Pink) ($\alpha = 4.5$): small clusters that form and dissolve rapidly;  Turbulent (Red) ($\alpha = 7.5$): chaotic, swirling motion with frequent rod collisions; Flocking (Yellow): swarms of rods merge to form larger flocks moving collectively for $\alpha=11.6$ and $\alpha = 15.1$; Large clusters (Purple): large clusters form with limited mobility in the core for $\alpha = 7.5$, $\alpha= 11.6$ and $\alpha = 15.1$; Phase separation (Gray): dense clusters coexisting with dilute areas for $\alpha = 4.5$;  Jammed (Brown): complete suppression of rod mobility.}
	\label{fig:4}
\end{figure}

Shorter rods ($\alpha = 4.5$) at low area fractions ($\phi \leq 0.08$) exhibit an isotropic state characterized by random propulsion without alignment or clustering (movie S1). This disordered state persists until $\phi \approx 0.12$, where transient clusters begin to emerge (Fig. \ref{fig:4}). As $\phi$ increases to $0.27$, these transient clusters grow in size, though they remain dynamically unstable as evidenced by the temporal variation of overlaid cluster sizes as shown in movie S8. The transient nature of these clusters is attributed to the relatively high rotational diffusion coefficient of shorter rods (see fig. S4), which disrupts persistent alignment. At higher densities ($\phi \approx 0.6$), the system undergoes phase separation, marked by the coexistence of dense, rod-rich clusters and dilute, gas-like regions (Fig. \ref{fig:3} and movie S6). These clusters remain dynamic and mobile but do not exhibit coherent collective motion or global polarity.

Compared to shorter rods ($\alpha = 4.5$), rods with $\alpha = 7.5$ display a richer and more ordered sequence of collective states (see Fig.~\ref{fig:1}C–F and Fig.~\ref{fig:3}, see movie S1), driven by enhanced shape-induced alignment and stronger hydrodynamic interactions. At low area fractions, both systems remain in an isotropic state, but for $\alpha = 7.5$, the transition to collective motion occurs earlier, at $\phi \approx 0.07$, marked by the emergence of stable, polar-aligned clusters. As the density increases to $\phi = 0.23$, these clusters evolve into a coherent swarming state, in contrast to the $\alpha = 4.5$ system, which continues to form only transient, disordered clusters.
At $\phi = 0.39$, the $\alpha = 7.5$ rods enter a turbulent phase characterized by chaotic swirling motion, while the shorter rods still exhibit transient clustering without sustained coherence (movie S5). At higher area fractions ($\phi = 0.72$), steric interactions suppress turbulence in the $\alpha = 7.5$ system, resulting in large, stable clusters and eventual jamming near $\phi \approx 0.9$. In contrast, the $\alpha = 4.5$ system undergoes phase separation at high density ($\phi = 0.6$) but fails to develop local polarity or sustained collective motion.

Rods with a high aspect ratio ($\alpha = 15.1$) display isotropic motion only at low concentrations (see movie S1), up to $\phi \approx 0.05$, lower than for smaller aspect ratios, indicating an earlier onset of collective motion due to enhanced shape-induced anisotropic alignment. Above $\phi = 0.06$ rods form small aligned clusters that evolve into persistent swarms at intermediate area fractions ($\phi = 0.26$). Increasing the area fraction further to $\phi=0.38$ leads to the merging of swarms into larger structures that resemble flocks (Fig.~\ref{fig:3} and movie S9). These flocks move slower and have lower orientational order compared to swarms, but still exhibit cooperative motion. Unlike the turbulence observed for $\alpha = 7.5$ at $\phi = 0.39$, turbulence is suppressed for highly anisotropic rods at $\phi=0.38$ despite exhibiting similar flow fields (see fig. S6 and S7), as steric interactions dominate over hydrodynamic forces. This leads to a flocking state with reduced chaotic motion, consistent with previous simulation studies showing that high anisotropy suppresses turbulence\cite{zantopEmergentCollectiveDynamics2022}. At higher area fractions ($\phi = 0.65$), dense clusters merge into larger structures exhibiting slow, coordinated motion, contrasting with the dynamic clustering seen in rods with lower aspect ratios. Finally, at $\phi \approx 0.9$, the system becomes jammed.
The state behavior of rods with $\alpha = 11.6$ is strikingly similar to that of higher aspect ratio rods ($\alpha = 15.1$), except that the area fractions corresponding to each state are slightly shifted to the right, as shown in Fig.~\ref{fig:4}. 

As the aspect ratio ($\alpha$) of the rods increases, the range of area fractions over which coherent motion is observed also expands. For instance, rods with an aspect ratio of $\alpha = 15.1$ begin to exhibit collective motion at a lower $\phi$ compared to shorter rods with $\alpha = 4.5$. The increased tendency for alignment in more anisotropic rods, resulting from shape-induced alignment, combined with reduced rotational diffusion (fig. S4 and S17), facilitates the onset of collective behavior at lower concentrations. These findings align with theoretical and simulation predictions that higher aspect ratios promote clustering and coherent motion\cite{zantopEmergentCollectiveDynamics2022,grossmann2020particle}. Furthermore, a higher aspect ratio reinforces polar order, leading to more stable and long-lived collective motion, consistent with bacterial swarming behavior\cite{beerPhaseDiagramBacterial2020}. 
The differences in collective behavior between different aspect ratios contribute to distinct non-monotonic trends observed in the mean speed as a function of concentration (see fig. S5).

To decouple the roles of steric and hydrodynamic interactions in the emergent collective dynamics of self-propelled rods (see SI sections 2.3-2.5), we performed active Brownian dynamics simulations in which hydrodynamic interactions were excluded \cite{yangSwarmBehaviorSelfpropelled2010,vutukuriDynamicSelforganizationSidepropelling2016}. In this model, rods are represented as bead chains with propulsion applied only at the head, similar to the experimental configuration (see Methods). The resulting collective behavior (see movie S10) and phase transitions arise solely from repulsive steric interactions, the number of rods, and rotational diffusion.
In simulations, we constructed the state diagram based on a quantitative analysis of the cluster size distribution (fig. S20, S21). The absence of a turbulent state at $\alpha=7.5$ in these simulations underscores the crucial role of hydrodynamic interactions, present in the experiments, in enabling the emergence of turbulence. Overall, the simulation results are consistent with experimental observations, showing that higher anisotropy promotes the onset of collective behavior at lower area fractions (fig. S19). Moreover, the simulations exhibit similar trends in the cluster size distribution, number fluctuations, and the velocity autocorrelation function (figs. S20, S22, and S23), in agreement with the experimental results (Fig. \ref{fig:2}).
\\

{\noindent {\bf {Conclusions}}\label{sec13}}\\
Our combined experimental and simulation study demonstrates how shape-induced alignment, particle concentration, and hydrodynamic interactions govern the self-organization of active rods. By systematically varying the area fraction $\phi$ and the aspect ratio of the rods, we constructed a state diagram and identified distinct collective states, ranging from isotropic motion to swarming, active turbulence, transient clusters, flocking, large clusters, and jamming. Our results reveal that hydrodynamic interactions, including fluid-mediated torques and flow fields, combine with shape-induced alignment to play a central role in driving the system into the active turbulence state for intermediate aspect ratio rods ($\alpha=7.5$). Brownian dynamics simulations without hydrodynamics reproduce clustering and swarming behavior, but fail to capture turbulence, underscoring the essential role of hydrodynamic torques in dynamically reorienting rods and generating local vorticity and rotational stress necessary for the emergence of active turbulence.

Comparative analysis across different aspect ratios of active rods reveals that highly anisotropic rods ($\alpha = 11.6, 15.1$), characterized by stronger alignment, exhibit swarming motion at lower area fractions, suppressed active turbulence, and enhanced flocking. In contrast, shorter rods ($\alpha = 4.5$) display transient clustering and motility-induced phase separation, resulting from higher rotational diffusion and weaker alignment interactions. We further quantify non-equilibrium behavior via giant number fluctuation analysis, which reveals anomalous scaling in the turbulent state, and velocity autocorrelation functions (VACF), which show distinct decay profiles across different states. The energy spectra in turbulent phases show two characteristic scaling regimes, consistent with active turbulence \cite{alertActiveTurbulence2022}. 
This work provides a framework for understanding shape-dependent collective dynamics in active matter and stimulates the development of more realistic theoretical and simulation models of polar active rods, with implications for programmable swarming, adaptive transport, and biomimetic materials. To contextualize our results, we compare them to bacterial turbulence in \textit{E. coli} and \textit{B. subtilis}, where wild-type strains with intermediate aspect ratios ($\alpha \approx 5.6-6.3$) exhibit turbulence \cite{wensinkMesoscaleTurbulenceLiving2012,pengImagingEmergenceBacterial2021}, while elongated mutants ($\alpha \geq 13-25$) do not \cite{nishiguchiLongrangeNematicOrder2017,beerPhaseDiagramBacterial2020}. Our study offers a physical rationale for this long-observed but previously unexplained biological trend, suggesting that natural swimmers may have evolved aspect ratios that optimize collective transport and mixing.\\




{\noindent \bf{ACKNOWLEDGMENTS}}\\
The authors thank the Nanobiophysics and Physics of Complex Fluids groups (University of Twente, The Netherlands) for providing access to their laboratory facilities. The authors thank Stijn van der Ham for COMSOL simulations and fruitful discussions, Andeas Z\"ottl and Shravya Narendula for useful discussions, Ineke Punt and Tjalling Ritsema for SEM measurements, and SURF (www.surf.nl) for support in using the National Supercomputer Snellius. We also thank the reviewers for their constructive suggestions.
{\noindent \bf{Funding:}} H.R.V. acknowledges funding from The Netherlands Organization for Scientific Research (NWO-M1-OCENW.M.21.309, \&
OCENW.XS23.4.115) and the European Research Council (ERC Consolidator Grant no.
101171050-SynthAct3D).
{\noindent \bf{Author contributions:}}
H. R. V. conceived,  supervised the project, and acquired funding. H. R. V., Y. S., and A. N. S. designed the research. Y. S. synthesized the particles and performed all experiments. A. N. S. performed the simulations. All authors participated in discussions, data analysis, and were involved in writing, reviewing, and editing the manuscript.
{\noindent \bf{Competing interest:}}
The authors declare no competing interests. 
{\noindent \bf{Data and materials availability:}}
All data reported in the main text and Supplementary Information, as well as executable scripts for the numerical simulations, are available at~\cite{4tudata}. Details for the synthesis and characterization of all particles and materials used in this study are provided in the Supplementary Materials.\\

{\noindent\bf{SUPPLEMENTARY MATERIALS}}\\
Materials and Methods\\
Supplementary Text\\
Figs. S1 to S24\\
Table S1 to S3\\
References (42-52)\\
Movies S1 to S10

\clearpage
\newpage

\setcounter{equation}{0}
\setcounter{figure}{0}
\setcounter{table}{0}
\setcounter{page}{1}
\setcounter{section}{0}

\makeatletter
\renewcommand{\theequation}{S\arabic{equation}}
\renewcommand{\thefigure}{S\arabic{figure}}
\renewcommand{\thetable}{S\arabic{table}}
\renewcommand{\thesection}{S\arabic{section}}
\makeatother

\begin{center}
    \includegraphics[width=0.3\textwidth]{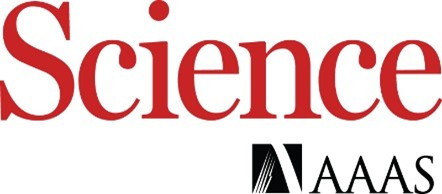} 
\end{center}

\vspace{0.5cm}

\begin{center}
    {\Large \textbf{Supplementary Materials for}}
    
    \vspace{0.4cm}
    
    {\Large \textbf{Shape Anisotropy Governs Organization of Active Rods: Swarming, Turbulence, Flocking, and Jamming}}
    
    \vspace{0.4cm}
    
    {\large Yogesh Shelke, Anpuj Nair S, and Hanumantha Rao Vutukuri}
    
    \vspace{0.4cm}
    
    Correspondence to: h.r.vutukuri@utwente.nl
    
\end{center}

\vspace{0.5cm}

\noindent \textbf{This PDF file includes:}
\begin{itemize}
    \item Materials and Methods
    \item Supplementary Text
    \item Figs. S1 to S24
    \item Tables S1 to S3
    \item Captions for Movies S1 to S10
\end{itemize}

\vspace{0.5cm}

\noindent \textbf{Other Supplementary Materials for this manuscript include the following:}
\begin{itemize}
    \item Movies S1 to S10
\end{itemize}

\newpage

\section{Materials and Methods}

\subsection{Materials} 
 Polyvinylpyrrolidone (PVP, average molecular weight Mn = 40,000, Sigma-Aldrich), 1-pentanol ($\geq$99$\%$, Sigma-Aldrich), ethanol, ammonium hydroxide (28--30\%, Sigma-Aldrich), sodium citrate dihydrate (99$\%$, Sigma-Aldrich), tetraethyl orthosilicate (TEOS, $\geq$ 98$\%$, Sigma-Aldrich), methanol ,  acetonitrile (ACN, $\geq$ 99.5$\%$, Sigma-Aldrich), dodecylamine (DDA, $\geq$ 98$\%$, Sigma-Aldrich), titanium isopropoxide (TTIP, 97$\%$, Sigma-Aldrich), hydroquinone (HQ, ReagentPlus®, $\geq$99$\%$, Sigma-Aldrich), p-benzoquinone (BQ, reagent grade, $\geq$98$\%$, Sigma-Aldrich), imidazole (ACS reagent, $\geq$99$\%$, Sigma-Aldrich), Milli-Q (Milli-Q\textsuperscript{\textregistered} EQ 7000) water with resistivity 18.2 M$\Omega \cdot$cm was used for all experiments. All chemicals were used as received from the supplier.

\subsection{Synthesis of silica rods} 

Rods with a length of $2 \pm 0.1$ $\mu$m and a diameter of $0.45 \pm 0.05$ $\mu$m ($\alpha = 4.5\pm 0.34$) were prepared using the method described in \cite{mu2022light, kuijk2011synthesis}. Briefly, in a 50 mL centrifuge tube, 4 g of PVP  was completely dissolved in 40 mL of 1-pentanol by sonication for 2 hours (Branson 1510). Once the PVP had completely dissolved, 4 mL of absolute ethanol, 1.12 mL of ultrapure water and 0.264 mL of 0.18 M sodium citrate dihydrate solution were added to the pentanol-pvp solution. The tube was then shaken by hand to mix the contents for 2 minutes. Then, 0.9 mL of ammonium hydroxide was added, the bottle was shaken again, and 0.4 mL of TEOS was added to the mixture. After shaking for another 2 minutes, the tube was left to rest overnight to allow the reaction to proceed for 15–16 hours. Following the reaction, the mixture was centrifuged at 3500 rpm (ThermoFisher Scientific Megafuge 8 Small Benchtop Centrifuge Series) for 10 minutes. The supernatant was removed, and the sedimented particles were redispersed in methanol. This centrifugation procedure was repeated at 3000 rpm for 10 minutes, again repeated with methanol, and the particles were stored in 40 mL methanol. Longer rods were synthesized by sequential post-growth additions of TEOS. Each addition consisted of 0.4 mL of TEOS added to a different batch with a similar reaction mixture at 8–12 hour intervals. A single addition of TEOS, introduced after 8 hours, produced rods with a length of $5.35 \pm 0.55$~$\mu$m and a diameter of $0.71 \pm 0.14$~$\mu$m ($\alpha = 7.5 \pm 1.8$). A second addition, 8–12 hours later, resulted in rods with a length of $8.15 \pm 0.4$~$\mu$m and a diameter of $0.7 \pm 0.12$~$\mu$m ($\alpha = 11.6 \pm 1.9$). One further TEOS addition, also spaced 8–12 hours apart, produced the longest rods with a length of $10.9 \pm 0.72$~$\mu$m and a diameter of $0.72 \pm 0.17$ ~$\mu$m ($\alpha = 15.1 \pm 2.15$). After the desired growth was achieved, the rods were cleaned with methanol using the same procedure as mentioned above. The shape and size distribution of the synthesized rods were characterized by scanning electron microscopy (SEM). Representative SEM images are shown in Figure \ref{fig:SEM}.

\subsection{Synthesis of \(\mathrm{SiO}_{2}\)-\(\mathrm{TiO}_{2}\) rods}

To grow $\mathrm{TiO_2}$ spheres on one end of the rods, we followed the procedure reported in \cite{mu2022light}. Typically, a suspension \(\sim 2\%\) (w/v) of silica rods in 40 mL of methanol was transferred to a 100 mL Pyrex bottle. Then, 20 mL of ACN was added. Next, 62 $\mu$L of deionized water and 128~$\mu$L of DDA were added to the glass bottle. The mixture was sonicated for 2 minutes to ensure the dispersion of the \(\mathrm{SiO}_{2}\) rods and then stirred at 300 rpm for an additional 8 minutes. Following this, 258 $\mu$L of TTIP was added to the mixture, and the reaction was stirred at 100 rpm for 6 hours at 20~$^\circ$C. Finally, the product was washed with ethanol under centrifugation for 2 minutes at 4200 rpm to remove the $\mathrm{TiO_2}$ spheres formed by secondary nucleation. The synthesized (\(\mathrm{SiO}_{2}\)--\(\mathrm{TiO}_{2}\)) particles were then immersed in a 1~mL of 0.8 $\text{mM}$ D5 dye ethanolic solution for 15 hours to complete the dye absorption on the $\mathrm{TiO_2}$ surface. Then, excess supernatant was removed, and the rods were again cleaned with water by centrifugation for 2 minutes at 2000 rpm. 

\subsection{Synthesis of bare silica tracers}
Silica tracers of size $\sim700$~nm were synthesized via the St\"{o}ber method, as described in Ref.\cite{carrasco2025characterization}. Briefly, a mixed solution containing 69 mL of isopropanol, 2.9 mL of 6 M ammonia (NH$_3$), and 7.43 mL of deionized water was prepared in a reaction vessel. The solution was stirred at 200 rpm for 15 minutes at room temperature. Subsequently, 4 mL of tetraethyl orthosilicate (TEOS) was added dropwise at a rate of approximately 2 drops/s using a micropipette while maintaining stirring. The reaction vial was then covered with paraffin film, and the mixture was allowed to react for 18 hours at room temperature. Following the reaction, the resulting suspension of bare silica tracers was cleaned by centrifugation once with ethanol and finally twice with water at 800 rpm for 20 min to a final volume of 200 mL.


For the synthesis of larger $\sim950$~nm sized silica tracers, we used the optimized St"{o}ber method described in \cite{zhaoOnestepSynthesisMicronsized2025}. Two solutions were first prepared separately: A (4 ml of TEOS in 30 ml of ethanol) and B (12 ml of NH$_3$.H$_2$O in 50ml of ethanol), and each was magnetically stirred for 10 min to mix the solutions thoroughly. Solution A was then added dropwise to the beaker containing solution B using a micro syringe pump at a rate of 20 mL/h. Once the addition was complete, the entire mixture was allowed to react with continuous stirring for 24 h. Finally, tracers were cleaned by centrifugation, once with ethanol and then twice with water, at 800 rpm for 10 minutes.

\subsection{pNIPAM coating of silica tracers}

To disentangle phoretic flows from osmotic flows, we coated tracer particles with a pNIPAM brush layer of thickness $\sim 45$~nm, following the method described in Ref.\cite{carrasco2025characterization}. This process involves functionalizing the silica surface with TPM as a grafting agent, followed by \textit{in situ} polymerization with pNIPAM. To functionalize the surface, 200 mL of the synthesized silica suspension (in ethanol) was added to a Pyrex bottle. Then, 5 mL of 3-(trimethoxysilyl) propyl methacrylate (TPM) was added, and the mixture was stirred at 200 rpm for 12 hours at room temperature. To promote covalent Si--O--Si bonding of TPM to the silica surface, after 12 hrs suspension was transferred to a 300 mL three-neck round-bottom flask equipped with a reflux condenser and subsequently was heated to reflux at $\sim 80^\circ$C for 1 hour while stirring at 200 rpm in a temperature-controlled water bath. Finally, the TPM-functionalized silica spheres were purified twice with centrifugation at 800 rpm for 10 minutes and redispersed in ethanol.

To coat the particles with pNIPAM via \textit{in situ} polymerization, an aqueous solution consisting of 1.132 g of N-isopropylacrylamide (NIPAM) and 0.078 g of N,N$'$-methylene bisacrylamide (BA) in 100 mL of deionized water was prepared in a 150 mL three-neck round-bottom flask equipped with a reflux condenser and immersed in a temperature-controlled oil bath. The solution was stirred at 300 rpm for 10 minutes at 40$^\circ$C. Then, 2 mL of the concentrated TPM-functionalized silica suspension was added to the mixture, and stirring continued at 70$^\circ$C for 1 hour. Subsequently, 50 $\mu$L of sodium persulfate (KPS) initiator (0.04 g/mL) was quickly added while maintaining a continuous N$_2$ gas flow. The reaction was stirred for 4 hours at 70$^\circ$C, during which the solution typically became turbid within $\sim$15 minutes, indicating polymerization. After cooling the reaction mixture to room temperature, the pNIPAM-coated silica particles were purified, and secondary-formed pNIPAM particles were removed by at least five cycles of centrifugation at 800 rpm for 30 minutes and redispersion in water.

\subsection{Sample preparation for collective dynamics study of active rods}
The synthesized (\(\mathrm{SiO}_{2}\)--\(\mathrm{TiO}_{2}\)) rods were suspended in fuel solution to enable light-activated propulsion. Specifically, 6~\(\mu\)L of a rod dispersion at the desired concentration was mixed with 10~\(\mu\)L of 40~mM HQ, 2~\(\mu\)L of 20~mM BQ, and 2~\(\mu\)L of 2~mM imidazole. The resulting mixture was then sonicated (Branson 1510), loaded into a rectangular capillary, and sealed with UV-curable glue (Norland Optical Adhesive 68). The sealed capillary was placed upside down on the microscope stage, allowing the rods to settle onto the bottom surface of the capillary by gravity. After sedimentation, the rods were activated using a fluorescence illumination system integrated into the microscope, equipped with a green excitation filter with wavelength ($\lambda \sim$ 532\text{–}552\,\text{nm}). The light intensity was maintained at \(33 \pm 5\)~mW/cm\(^2\), as measured using power meter (Coherent PowerMax PM10). Illumination conditions were kept constant across all experiments to ensure uniform activation of rods with different aspect ratios. Imaging was conducted on a Nikon Eclipse TE2000-U inverted microscope equipped with a Basler acA4112-30\(\mu\)m CMOS camera. Objective lenses of 40\(\times\) (air, N.A.~\(= 0.75\)), 60\(\times\) (oil, N.A.~\(= 1.40\)), and 100\(\times\) (oil, N.A.~\(= 1.30\)) were used depending on the resolution required. Videos were recorded at 20-30 frames per second under bright-field and/or fluorescence imaging modes.

\subsection{Flow fields around immobilized  \(\mathrm{SiO}_{2}\)-\(\mathrm{TiO}_{2}\) rods}
To visualize the fluid flow around active rods, a drop of a dilute concentration of active rods was spread on a plasma-cleaned glass substrate. After drying at room temperature, the active rods were immobilized on the substrate. A 200~$\mu$L pipette tip, cut into a small well, was then glued on top of the substrate. The well was filled with a solution containing 10~$\mu$L of 40~$\text{mM}$ HQ, 2~$\mu$L of 20~$\text{mM}$ BQ, 2~$\mu$L of 2~$\text{mM}$ imidazole, and 6~$\mu$L of 700~nm spherical silica tracer particles as passive tracers.
Under green light illumination, the response of tracer particles to the immobilized active rods was recorded at 50 or 100~fps. The light illumination was periodically switched on and off to modulate the rods’ activity, enabling both attractive and repulsive flow fields to be resolved~\cite{carrasco2025characterization}.
During these on–off cycles, the displacement of 700~nm silica tracer particles was tracked over time. To compute the net velocity fields and streamlines generated by the fluid around the rods, ensemble-averaged particle image velocimetry (PIV) was applied\cite{thielicke2021particle}.

 {To ensure robustness of the flow characterization, we employed tracer particles of different sizes and materials, including bare and pNIPAM-coated silica (SiO$_2$) spheres and TPM spheres, with diameters ranging from 0.54~$\mu$m to 1.04~$\mu$m. The silica tracer particles were synthesized using a modified Stöber method, the TPM spheres were prepared following emulsion polymerization protocols, and the pNIPAM shell on the silica spheres was grown using the protocol in \cite{carrasco2025characterization}. The surface charges of these tracers, quantified by their zeta potentials in the reaction medium, are summarized in Table~\ref{tab:zeta}. 

In order to obtain the decay of flow field velocity with distance around an immobilized rod, we first tracked the tracers, radially binned the data, and then averaged the instantaneous tracer velocities for only the tracers in the plane of the rod. We only consider tracers near the front of the head because the flows are strongest there. Since the flow is radially outward from the head here, we only consider the radial component of tracer velocity, $v_r$. The excluded angular component of velocity is mostly Brownian noise.

\subsection{Rod tracking, orientation extraction, and flow field analysis}

To track the positions of rod heads, we used the TrackMate plugin in ImageJ \cite{ershovTrackMate7Integrating2022}, which enabled automated detection and linking of rod head positions across sequential image frames. This approach was effective for rod aspect ratios $\alpha = 4.5$, $7.5$, and $15.1$ and for area fractions up to $\phi \approx 0.8$, resulting in continuous trajectories that captured the translational motion of individual rods. To determine rod orientations at lower area fractions ($\phi \leq 0.4$), images were first preprocessed by converting them from grayscale to binary using Otsu’s thresholding method. Morphological operations such as erosion, dilation, and particle filtering were then applied to remove background noise and isolate individual rod shapes. From the segmented binary images, rod positions and orientations were extracted using the thresholding method in TrackMate. To extract the flow field of the active rods, we employed particle image velocimetry (PIV)\cite{thielicke2021particle}. Each microscopy frame was divided into interrogation windows with 50$\%$ overlap, and cross-correlation between successive frames was used to compute local displacement vectors. All extracted trajectory, orientation, and flow field data were post-processed using custom Python scripts to extract relevant parameters.

\subsection{Characterization} 

SiO$_2$–TiO$_2$ rods were characterized using scanning electron microscopy (SEM) imaging. Before SEM imaging, a 5\,nm thin platinum/palladium layer was deposited on the surface of the rods using a metal sputter coater (QUORUM Technologies, Q150TS E Plus). SEM imaging was performed using a JEOL JSM-6010A microscope operated at 5\,kV to accurately determine the geometry of the rod, including length, diameter, and surface features. The area fraction of rods in each experimental frame was calculated by analyzing a single frame taken just after light activation of rods. The number of rods was estimated by counting the identifiable rod heads in the field of view. To calculate the area of the rods, each rod was approximated as a cylinder, and its projected area was estimated using its measured length and diameter. The area fraction was then computed as the total projected area of all rods divided by the area of the imaged field of view.

\subsection{Calculation of cluster size distribution}

To quantify the degree of clustering across different area fractions, we computed the probability distribution of cluster sizes, $P(n)$, where $n$ denotes the number of rods in a cluster. In experiments, we extracted the $(x, y)$ positions of the TiO$_2$ head of each rod from bright field and fluorescence images. At high $\phi \geq 0.4$, we were only able to detect the heads of the rods, so we used head positions in all concentrations. Clusters were identified using the DBSCAN algorithm from the scikit-learn library\cite{scikit-learn}, with a radial distance cutoff of $2\,\mu$m and \texttt{min\_samples} = 2. These parameters allowed for the reliable detection of clusters in all densities. This analysis was repeated over time to ensure sufficient statistics. In simulations, we used the same DBSCAN algorithm but applied it to the positions of all beads composing each rod. Two rods are considered part of a cluster if any of one rod's beads fall within a distance of $epsilon = 1.1\,\sigma$ of the other rod’s beads. The value of \texttt{min\_samples} was again set to 2. For each simulation, we calculated $P(n)$ by counting the number of clusters of size n over large times ($t>250\tau$) after every $10\tau$. 

\subsection{Velocity autocorrelation function}
In experiments, the velocity autocorrelation function (VACF) was calculated using Eq. \ref{eq:VACf} from trajectories of the rod heads extracted using the Fiji plugin TrackMate \cite{ershovTrackMate7Integrating2022}. VACF is a measure of persistence of motion that quantifies how correlated the velocity of a rod is with its own velocity after a time interval $\Delta t$.

\begin{equation}
    C_{v}(\Delta t) = \biggl \langle \frac{\langle \mathbf{v}_i(t+\Delta t) \cdot \mathbf{v}_i(t) \rangle_{t}}{\langle \mathbf{v}_i(t) \cdot \mathbf{v}_i(t) \rangle_{t}} \biggr \rangle _i
    \label{eq:VACf}
\end{equation}
where $\mathbf{v}_i(t)$ is the velocity vector of the i-th rod at time t, $\langle\cdot\rangle_i$ is the average over all the rods i, and $\langle\cdot\rangle_t$ is the average over all time intervals. In our experiments, recorded at 20 frames per second, the instantaneous velocity captures the Brownian motion and gets decorrelated. Hence, we use the average velocity over 0.5 s to calculate the VACF in Fig. 2.

\subsection{Spatial velocity correlation function}

The spatial velocity correlation function is defined as:

\begin{equation}
    C_s(\Delta r)= \biggl \langle \frac{\langle \mathbf{v}(\mathbf{r+\Delta r},t) .\mathbf{v}(\mathbf{r},t) \rangle_\mathbf{r}}{\langle \mathbf{v}(\mathbf{r},t) .\mathbf{v}(\mathbf{r},t) \rangle_\mathbf{r}}\biggr \rangle _t
\end{equation}
where $\mathbf{v}(\mathbf{r})$ is the velocity field at $\mathbf{r}$ obtained from PIV, $\mathbf{r+\Delta r}$ represents all points at a distance $\Delta r$ from $\mathbf{r}$, $\langle\cdot\rangle_\mathbf{r}$ is the average over all such points and all $\mathbf{r}$. We use a bin size of 2.2 $\mu$m to bin points based on radial distance. This function essentially describes how the correlation between the velocities of a pair of rods changes with the distance between them. When vortices are present in the system, this function decays to negative values, and the minima correspond to the typical size of a vortex. In the case of mesoscale turbulence in bacteria\cite{wensinkMesoscaleTurbulenceLiving2012}, a spatial correlation similar to that in our synthetic rods was observed.

\subsection{Calculation of translational and rotational mean squared displacement}
To quantify anisotropic diffusion of rods in experiments, we calculated the mean squared displacement (MSD) in the body frame of each rod. Frame-to-frame displacements \((\Delta x, \Delta y)\) in the laboratory frame were projected into the rod’s body frame using its orientation angle \(\theta\), according to the rotation:

\begin{equation}
    \begin{pmatrix}
        \Delta r_\parallel \\
        \Delta r_\perp
    \end{pmatrix}
    =
    \begin{pmatrix}
        \cos\theta & \sin\theta \\
        -\sin\theta & \cos\theta
    \end{pmatrix}
    \begin{pmatrix}
        \Delta x \\
        \Delta y
    \end{pmatrix}
\end{equation}
where \(\Delta r_\parallel\) and \(\Delta r_\perp\) are the displacements along and perpendicular to the rod's long axis, respectively. The MSDs in each direction were computed as:

\begin{align}
    \langle \Delta r_\parallel^2(t) \rangle &= \langle [r_\parallel(t)- r_\parallel(0)]^2 \rangle \\
    \langle \Delta r_\perp^2(t) \rangle &= \langle [r_\perp(t)- r_\perp(0)]^2 \rangle
\end{align}
The translational diffusion coefficients, \(D_\parallel\) and \(D_\perp\), were obtained by fitting the linear regime of the MSDs to \(v^2t^2+2D_{\parallel} t\) and \(2D_{\perp} t\), where \(t\) is the lag time. To measure rotational diffusion, the rod orientation \(\theta(t)\) was tracked over time, and the mean squared angular displacement was computed as:

\begin{equation}
    \langle \Delta \theta^2(t) \rangle = \langle [\theta(t) - \theta(0)]^2 \rangle
\end{equation}
The rotational diffusion coefficient, \(D_R\), was extracted by fitting this curve to the relation:

\begin{equation}
    \langle \Delta \theta^2(t) \rangle = 2 D_R t
\end{equation}
In experiments, the centroid and orientation of each rod were extracted using Trackmate \cite{ershovTrackMate7Integrating2022} from the positions of the TiO$_2$ head and SiO$_2$ tail using bright-field and/or fluorescence microscopy, and analyzed with custom image processing Python scripts. 

\YS{
\subsection{Energy Spectrum for active turbulence}

The calculation of the energy spectrum is performed to characterize turbulence by analyzing the velocity field based on the procedure in \cite{liuDensityFluctuationsEnergy2021}. The energy spectrum, E(k), quantifies the distribution of kinetic energy across different length scales, where k is the wavenumber. In order to reliably obtain the spectrum for a wide range of k, we perform experiments with a larger field of view (325~$\mu$m $\times$ 325~$\mu$m) in the turbulent phase ($\alpha=7.5,\phi\approx0.4$). The velocity field $\mathbf{v(r)}$ is obtained in a regular grid with spacing 1.73~$\mu$m  at different times using Particle Image Velocimetry (PIV) in PIVlab \cite{thielicke2021particle}. For PIV, we use 3 passes with $50\%$ overlap, followed by filtering erroneous velocities with a low contrast threshold. Smoothing was avoided to prevent changes to the actual velocity fields, and interpolation was used for $\phi\approx0.4$.

The Fourier transform of the velocity field, $\mathbf{\tilde{v}(k)}$, is calculated as
$$ \tilde{\mathbf{v}}(\mathbf{k}) = \int_{\mathbb{R}^2} \mathbf{v(r)} e^{-i \mathbf{k} \cdot \mathbf{r}} d\mathbf{r} $$
where $\mathbf{k}$ is the wavevector. 

Since we use a fast Fourier transform (FFT) instead of a continuous Fourier transform to convert $\mathbf{v}(x,y)$ to $\mathbf{\tilde{v}}(k_x,k_y)$, we have to multiply by the square of the spacing in the velocity field (1.73~$\mu$m). The kinetic energy density at a point $\mathbf{k}$ in k-space, $\tilde{E}(\mathbf{k})$, is calculated as:
$$ \tilde{E}(\mathbf{k}) = \frac{1}{2A} \left| \tilde{\mathbf{v}}(\mathbf{k}, t) \right|^{2} $$

Now, to obtain the energy spectrum, this $\tilde{E}(\mathbf{k})$ is radially binned in k-space and averaged. 

$$E(k) = 2\pi k \left< \tilde{E}(k,\theta) \right>_{\theta} $$

The final energy spectrum, $E(k)$, where $k=|\mathbf{k}|$, is obtained by averaging the energy over time and over multiple experimental runs. This method of calculating the energy spectrum ensures that $\langle|\mathbf{v(r)}|^2\rangle/2 = \int_0^{\infty} E(k) dk$.

}

\subsection{Simulation details}

We model synthetic self-propelled colloidal rods using Brownian dynamics simulations of chains of $N$ beads with size $\sigma$ connected in sequence with strong harmonic angles and bond potentials, ensuring the rods are rigid with the propulsion force acting only on the head bead. To mimic the head-tail asymmetry observed in our experimental system, we apply the active force only on the head bead, in contrast to previous simulations using a bead chain model\cite{yangSwarmBehaviorSelfpropelled2010}. For $\alpha=7$, the number of rods was varied from 470 rods for the isotropic state to 11649 rods for the jammed state at regular intervals to get 20 different area fractions.

The simulations were performed in LAMMPS\cite{LAMMPS} (version 2Aug2023) with Lennard-Jones units. This sets the length units as the particle diameter $\sigma$, energy units as $k_BT$ and time units as $\tau=\sqrt{\frac{m\sigma^2}{k_BT}}$ and the fundamental quantities $m,\sigma,\epsilon,k_B=1$. The overdamped Langevin equation was integrated with a timestep of $\Delta t = 10^{-5} \tau$. At every $10^4 \Delta t = 0.1 \tau$, the positions of the beads were stored. All simulations were run for a total time of $t=500\tau$ on the Snellius supercomputer using 128 cores. For Brownian dynamics simulations, the equation solved is the overdamped Langevin equation given by,
\begin{equation}
    \partial_t\boldsymbol{r_i} = \boldsymbol{\gamma}_t^{-1} \boldsymbol{F_i} + F_p \boldsymbol{p_i} + \sqrt{2} (\beta \boldsymbol{\gamma}_t)^{-1/2} \boldsymbol{\xi_t}
\end{equation}
where, $\boldsymbol{\gamma}_t$ is the friction tensor, $\boldsymbol{F_i}$ is the total conservative force acting on each bead, $F_p$ is the constant propulsion force which only acts on the head bead ($F_p = 0$ for other beads) along the direction $\boldsymbol{p_i} = (\cos\theta_i,\sin\theta_i)$, $\beta = (k_B T)^{-1}$ and $\boldsymbol{\xi_t}$ is a random variable with unit variance and zero mean. $\boldsymbol{F_i}$ consists of the forces experienced due to the harmonic angle and the bond potential between the beads and those due to interactions with other rods.

\begin{equation}
    \boldsymbol{F_i}=\boldsymbol{F_i^{WCA}}+\boldsymbol{F_i^{bond}}+\boldsymbol{F_i^{angle}}
\end{equation}
For the head bead, its orientation changes according to the equation:
\begin{equation}
    \partial_t\theta_i = \sqrt{2} (\beta \boldsymbol{\gamma}_r)^{-1/2} \xi_{\theta}
\end{equation}
where $\xi_\theta$ is a random variable with unit variance and zero mean, and $\boldsymbol{\gamma}_r$ is the rotational friction coefficient whose value changes with $\alpha$ based on the formula from \cite{tiradoComparisonTheoriesTranslational1984}.

\begin{equation}
    \frac{\pi \alpha^2}{3\gamma_R} = \ln{\alpha} - 0.662 + \frac{0.917}{\alpha} - \frac{0.05}{\alpha^2}
\end{equation}
The beads of different rods interact through a Weeks-Chandler-Andersen (WCA) potential, which is defined as:

\begin{equation}
    U_{WCA}= 4\epsilon\left[\left(\frac{\sigma}{r}\right)^{12} - \left(\frac{\sigma}{r}\right)^6 \right] + \epsilon \quad \text{for} \quad r<2^{1/6}\sigma
    \label{eq:wca}
\end{equation}

\begin{equation}
    \boldsymbol{F_i^{WCA}} = -\frac{\partial U_{WCA}(r)}{\partial r_j} \boldsymbol{\hat{r}}_{ij}
\end{equation}

\begin{equation}
    U_{bond}=k_{bond}(r-r_0)^2 \quad \quad U_{angle}=k_{angle}(\theta'-\theta'_0)^2
\end{equation}

\begin{equation}
    \boldsymbol{F_i^{bond}} = -\nabla_i U_{bond}(r) \quad \quad \boldsymbol{F_i^{angle}} = -\nabla_{ijk} U_{angle}(r)
\end{equation}

We set $r_0=\sigma$ and $\theta'_0=\pi$ to obtain the end-to-end length of the rod as $N\sigma$. Rod rigidity was enforced by using strong harmonic bond and angle potentials. Specifically, the bond and angle stiffness constants were set to \( k_{bond} = k_{angle} = 1000\,k_{\mathrm{B}}T/\sigma^2 \) for aspect ratios \( \alpha \leq 10 \), and increased to \( 2000\,k_{\mathrm{B}}T/\sigma^2 \) for \( \alpha > 10 \) to prevent bending due to asymmetric collision forces at high densities. For the repulsive WCA interactions between the rods, we use $\epsilon=100k_BT$. To achieve particles of diameter $\sigma$ in LAMMPS, we set the zero crossing distance of the Lennard-Jones potential to be $2^{-1/6}$, the cutoff at 1$\sigma$ followed by a potential shift to get the WCA potential in Eq. \ref{eq:wca}. The propulsion force was set to be $F_p=10k_BT$, the mass of each bead was set to 1, and the temperature was set to 0.1. The repulsion between the beads in a rod is turned off using the special bonds command in LAMMPS. LAMMPS dump files were parsed with the help of parsers in MDAnalysis\cite{michaud2011mdanalysis}. After turning off the propulsion, we ensure that the rod has the correct ratio between the parallel and perpendicular friction coefficients based on experimental values from \cite{yangBeadRodComparison2017}. The ratio between the translational drag coefficients varies from 1.18 for $\alpha=3$ rods to 1.43 for $\alpha=15$ rods.

We perform all our simulations in a box of size $300\sigma\times300\sigma$ with periodic boundary conditions. To check for system size effects, we varied the size of the simulation box from $300\sigma\times300\sigma$ to $600\sigma\times600\sigma$ and compared the cluster size distribution. The trend in the cluster size distribution was nearly identical. Hence, our $300\sigma\times300\sigma$ system does not have significant finite-size effects. Since the larger system can have bigger clusters as it has more particles, we observed a longer tail for the larger system. All simulations were initialized by placing rods on random sites in a lattice at the desired concentrations. The propulsion directions are also initially random. Even though the simulations are initialized on a lattice, after sufficient time, they become completely random and reach a steady state, where we can observe the collective behavior. The average cluster size vs time plateaus to a value dependent on concentration at $t>20\tau$. All the quantitative analyses performed here at large times were done for $t>250\tau$.

\newpage
\section{Supplementary Text}

\subsection{Mean squared displacements and gravitational height}

Diffusion coefficients \(D_{\parallel}\) and \(D_{\perp}\) were extracted by fitting the mean-squared displacements (MSD) in the body frame of the rods. We obtain $D_\parallel=0.5$ $\mu$m$^2$/s and $D_\perp=0.22$ $\mu$m$^2$/s for $\alpha=7.5$ rods at $\phi=0.01$. The rods exhibit ballistic motion along their long (parallel) axis, with an MSD vs time scaling exponent of 1.88, and sub-diffusive behavior in the perpendicular direction with an exponent of 1.19 (Fig. \ref{fig:exptmsdparperp}).
The nearly ballistic scaling in the parallel direction reflects persistent motion along the rod long axis.
Whereas passive rods are diffusive along both axes and yield lower values of translation diffusivities, $D_\parallel=0.13$ $\mu$m$^2$/s and $D_\perp=0.09$ $\mu$m$^2$/s, compared to active rods. This effect has also been observed in the case of side-propelled rods\cite{vutukuriDynamicSelforganizationSidepropelling2016}. We define the Péclet number as $\mathrm{Pe}=vL/D_\parallel = 39.6$ for $\alpha=7.5$ rods.

Rotational diffusion has a very drastic impact on the collective dynamics of active rods\cite{yangSwarmBehaviorSelfpropelled2010}. Theoretical studies have predicted that higher anisotropy leads to lower rotational diffusion\cite{tiradoComparisonTheoriesTranslational1984}. The rotational diffusion obtained from our experiments in Fig. \ref{fig:exptrotdiff} confirms these predictions. Based on the rotational diffusivity, we also calculate the rotational Péclet number as $\mathrm{Pe}_r=v/LD_r = 14.4$ for $\alpha=7.5$ rods. In the simulations, the orientations of each rod were calculated from the vector connecting the tail bead to the head bead. Subsequently, fitting the $\langle\Delta\theta^2\rangle$ vs t graph for $\alpha=7$ yields $D_R =0.0051$, and we obtain the rotational diffusion time $\tau_R=\frac{1}{D_R} = 19.6 \tau$. This timescale is similar to the timescale of decay of the velocity autocorrelation function as well as the time taken for the simulations to reach a steady state.

In our experiments, the rods are denser than the surrounding medium and therefore sediment under gravity onto the bottom glass substrate of the sample chamber.
We estimate the gravitational height, defined as the height at which the rod’s gravitational potential energy equals the thermal energy. For rods with aspect ratio $\alpha = 7.5$, this yields an average height of $109$~nm above the substrate. For longer rods ($\alpha = 15.1$), the gravitational height decreases further to $\sim$ 62~nm due to their increased mass. To further validate this estimate, we measured the translational diffusion coefficient of rods in their passive state and compared it with theoretical bulk values from Tirado \textit{et al.}\cite{tiradoComparisonTheoriesTranslational1984}, inferring the wall separation using the methodology described by Bitter \textit{et al.}\cite{bitterInterfacialConfinedColloidal2017}. The estimated average height from this diffusion-based analysis is $\sim 135$~nm, in good agreement with the gravitational height. We also note that under light illumination, and thus propulsion, the rods appear to move even closer to the wall, likely due to additional phoretic and osmotic interactions arising from chemical gradients near the substrate. However, direct measurement of the rod–wall separation in the active state is experimentally challenging and beyond the scope of this study. Therefore, the passive-state height estimates serve as an upper bound for the wall separation under propulsion.

\subsection{Mean speed of active rods}

Figure~\ref{fig:Velocity} shows the average speed of rods with different aspect ratios across various area fractions, revealing a non-monotonic trend. At low area fractions, the rod velocities remain comparable to those of isolated single rods. As the area fraction increases, the velocity increases and reaches a maximum at intermediate densities ($\phi \approx 0.3$–$0.4$). The exact mechanism underlying this velocity enhancement is not yet fully understood. However, we suspect that the increased propulsion speed at intermediate densities arises from a combination of hydrodynamic interactions and improved mixing of chemical fuel. A similar trend has been observed in bacterial systems, such as \textit{B. subtilis}, where hydrodynamic interactions are known to play a significant role~\cite{sokolovConcentrationDependenceCollective2007a}. At higher area fractions, steric interactions become dominant, reducing effective mobility and thereby decreasing the average velocity.

This trend of increased mean speed at intermediate concentrations can affect the state of collective dynamics, and the increase in speed is most pronounced for $\alpha=7.5$ rods, which have a turbulent phase where the chaotic collective motion enhances mixing and prevents steric jamming. In contrast, $\alpha=4.5$ rods are in the clustering phase at intermediate concentrations, and they are confined by steric interactions within dense clusters, which restricts their movement and lowers their average speed. Finally, $\alpha=15.1$ rods in the flocking phase show a slight increase in average speed caused by the synchronized, coordinated motion of the rods as they move together in flocks.

The magnitudes of the velocities also show an interesting trend with aspect ratio $\alpha$. Rods with $\alpha=15.1$ are only about half as fast as rods with $\alpha=7.5$ due to their mass being almost double, while the size of the head (and hence the propulsion force) is the same. Based on this justification, one would expect that rods with $\alpha=4.5$ would be much faster than $\alpha=7.5$, but this is not the case because the shorter rods are prone to more thermal noise. This results in their motion not being persistent and hence a lower mean speed.

{
\subsection{PIV measurements of flow fields}

We analyzed the flow fields using tracer particles of varying sizes, materials (silica, TPM), and surface chemistries (bare silica and polymer brush-coated silica). Specifically, we repeated the flow field analysis with different tracer sizes and materials, including 0.7~$\mu$m silica, 0.95~$\mu$m silica, and 0.75~$\mu$m TPM.

We find that both silica tracers of sizes 0.7~$\mu$m and 0.95~$\mu$m show similar flow fields (see Fig. \ref{fig:Flowfield_AR_7.5_and_15.1_silica_tpm} and Fig. \ref{fig:Flowfield_AR_7.5_bare_coated_1um}) and a $v_r(r) \sim 1/r^3$ decay for $\alpha = 7.5$ rods, as shown in Fig. \ref{fig:velocity_vs_distance_silica_1um_errorbar} and Fig. \ref{fig:velocity_vs_distance_silica_700nm_errorbar}. This scaling is consistent with theoretical predictions that account for both phoretic interactions and osmotic flows~\cite{andersonColloidTransportInterfacial1989}.
To assess the effect of tracer material, we quantitatively analyzed the velocity decay of TPM tracers in Fig.~\ref{fig:velocity_vs_distance_tpm}, which shows a decay exponent of $-2.82$, comparable to that observed with silica tracers.

Overall, all tracer particle results consistently show flow fields characteristic of pushers, confirming that the measured flows are robust and not tracer-specific with respect to size or material. We note that for tracer particles smaller than 0.54~$\mu$m, thermal noise dominates, making it difficult to reliably extract flow fields.

Finally, we compare the flow fields around rods with aspect ratios $\alpha = 7.5$ and $\alpha = 15.1$, using both silica and TPM tracer particles (Fig. \ref{fig:Flowfield_AR_7.5_and_15.1_silica_tpm}).
Our analysis shows that the radial flow velocity $v_r$ of tracer particles decays with distance $r$ as $v_r(r) \sim 1/r^3$, with comparable scaling for both aspect ratios. Moreover, the flow fields around the rods are qualitatively similar for both aspect ratios, and the extracted dipolar strengths are nearly identical \( (A \approx 37~\mu\mathrm{m}^4\,\mathrm{s}^{-1} ) \), as determined by fitting the radial velocity decay $v_r(r) = A/r^3$ (see Fig. \ref{fig:velocity_vs_distance_silica_700nm_errorbar}). Flow fields obtained via PIV analysis also reveal no significant differences between the two aspect ratios (see Fig. \ref{fig:Flowfield_AR_7.5_and_15.1_silica_tpm}), suggesting that the underlying propulsion mechanism is the same. This is expected, since the catalytic head, which generates the flow, is the same in both cases. The only difference is that the spatial extent of the flow becomes smaller relative to the total rod length in the case of $\alpha = 15.1$. 

Given that the propulsion mechanism and hydrodynamic character remain essentially unchanged, the emergence of turbulence at intermediate aspect ratios cannot be attributed to differences in driving forces. Instead, we identify the transition as governed by a subtle interplay between reorientation dynamics, steric alignment, and hydrodynamic torques. Only at intermediate aspect ratios do rods balance persistent propulsion, alignment interactions, sufficient reorientability, and hydrodynamic torques to generate spatially and temporally fluctuating flows, a hallmark of active turbulence.  This interpretation is further supported by the emergence of power-law scaling in the energy spectra, with exponents $0.8$ and $-3.2$. These scaling exponents are in an agreement with those observed in bacterial turbulence\cite{wensinkMesoscaleTurbulenceLiving2012,alertActiveTurbulence2022}, even though our system consists of minimal synthetic polar active rods without flagella, biochemical signaling, or adaptive behavior. At high aspect ratios ($\alpha \geq 11.6$), steric alignment dominates, leading to flocking on one hand, and on the other, suppressing fluctuations, which increases the likelihood of kinetic trapping.

\subsection{Decoupling phoretic effects from osmotic flows}

To disentangle phoretic flows from osmotic flows, we coated tracer particles with a pNIPAM brush layer of thickness $\sim$45~nm, following the method described in Ref.\cite{carrasco2025characterization}. These polymer brushes suppress phoretic motion by screening solute–surface interactions and reducing concentration gradients within the slip layer, thereby decreasing the effective phoretic mobility (zeta potential = –23.9 mV for 1.04 $\mu$m tracers). Under identical light intensity and fuel concentration, phoretic effects were largely suppressed, and tracer motion was dominated by osmotic outflow from the rod heads, with no inward tracer motion observed behind the heads of the rods. Tracer velocities in front of the rod head, $v_r(r)$, reflect the combined contributions of phoretic and osmotic flows for immobilized rods with $\alpha = 7.5$ (see Fig. \ref{fig:velocity_vs_distance_silica_1um_errorbar}). Bare tracers (orange) exhibit higher velocities dominated by phoretic effects, especially near the head, showing a consistent $1/r^3$ scaling. In contrast, brush-coated, sterically stabilized tracers (blue) exhibit much lower velocities near the head and a slower decay, confirming suppression of phoresis (see Fig. \ref{fig:velocity_vs_distance_silica_1um_errorbar}B).

Previous studies, which decouple phoretic and osmotic effects \cite{carrasco2025characterization,hardikarOsmoticPhoreticCompetition2024}, find that phoretic effects, which depend only on the chemical gradient, are symmetric and oppose osmotic flows. However, in our case, the underlying mechanism is complex, with both chemical and ionic gradients coming into play and resulting in asymmetric phoretic interactions with repulsive interactions in front of the rod and attraction behind the head of the rod.

\subsection{Interactions between rods}

To understand the nature of interactions between rods, we performed dilute-concentration experiments with mixtures of immobilized and freely moving rods. Upon light illumination, freely moving rods that approach an immobilized rod at an acute angle are attracted toward the immobilized rod’s head, remain behind the head for a short duration, and finally leave when the propulsion overcomes the attractions (see Figure~\ref{fig:rod-rod}). This behavior mirrors the phoretic attraction observed for passive tracer particles (Movie S2).
}

While tracer-based flow field measurements indicate a weak phoretic attraction just behind the head of the rod, our analysis of direct rod–rod interactions suggests that these are too weak to stabilize persistent pairs. For instance, when two oppositely moving rods meet head-on, the outward flows generated by each rod, both acting as pushers, result in an effective repulsion that is often strong enough to break the pair apart, as illustrated by two such events in Fig.~\ref{fig:rod-rod-free}. While we acknowledge that phoretic interactions may contribute, our experiments suggest that steric interactions and hydrodynamic flows between the rods are the primary drivers of collective behavior.

\subsection{Evolution of rod clusters over time}

To demonstrate that aligned rods synchronize both their orientation, we plot the local polar order parameter $\psi_{p}$ for a representative rod cluster (Fig. \ref{fig:Single_swarm}A and B).

\begin{equation}
\psi_{p} = \frac{1}{N} \left| \sum_{i=1}^{N} \frac{\mathbf{v}_i}{|\mathbf{v}_i|} \right|,
\label{eq:polar_order_parameter_orientaion}
\end{equation}
 where \(\psi_p \in [0,1]\) represents the degree of alignment, \(N\) is the number of rods in the cluster, and \(\mathbf{v}_i\) is the orientation of the \(i\)-th rod. \(\psi_p\) remained low (\(\psi_p \approx 0\)) for \(t < 4 s\), indicating no alignment. As the rods synchronized their orientation and velocity, \(\psi_p\) increased to 1 between \(t = 4 s\) and \(8.2 s\).

To test whether rods are geometrically interlocked in the swarming phase, we quantified the pairwise inter-rod distances between neighboring rods within a cluster for different cluster sizes (see Fig.~\ref{fig:fluct_comb}). We find that these distances fluctuate in an uncorrelated manner, indicating that rods are not locked in fixed positions. Instead, swarming emerges from shape-induced alignment and hydrodynamic interactions, rather than from geometric interlocking due to the rod caps.

\subsection{Tuning head size}

We synthesized rods whose head size is comparable to the tail and with $\alpha\approx7.5$. These new rods still exhibit swarming behavior at an area fraction $\phi = 0.12$, similar to our original rods with slightly larger heads (see Fig. \ref{fig:Small_head_rods}), under identical light intensity and fuel concentration. This demonstrates that swarming is not caused by the larger cap or head size. We also analyzed inter-particle distance fluctuations for these rods and observed slightly larger variations. This can be attributed to the reduced catalytic surface area of the smaller heads, which leads to weaker propulsion strength and, consequently, weaker effective hydrodynamic and phoretic interactions. In contrast, rods with larger heads generate higher propulsion due to their greater catalytic surface area exposed to the fuel, resulting in slightly stronger inter-rod attractions (see Fig. \ref{fig:fluct_smooth}).

\subsection{Simulation state diagram}

The simulation state diagram of self-propelled rods is constructed by systematically varying both the rod $\alpha$ and $\phi$, capturing how shape anisotropy and density affect the states of collective dynamics (see Fig. \ref{fig:simstatediag}). We performed Brownian dynamics simulations without hydrodynamic interactions to decouple the role of steric and hydrodynamic interactions. The different collective states were classified using the cluster size distribution (CSD), and further analyzed using velocity autocorrelation function (VACF), and giant number fluctuations (GNFs) (Fig. \ref{fig:numfluctsim} and  \ref{fig:vacfsim}).

All aspect ratios exhibit an isotropic state at the lowest area fractions with minimal clustering, indicating weak interactions. In the simulations, we used the cluster size distribution at large times to differentiate between the isotropic and swarming states. We designate a state as swarming when the maximum cluster size exceeds a threshold of 4\% of the total number of rods. As evident in Fig. \ref{fig:simstatediag}, this transition takes place at $\phi=0.21$ for $\alpha=5$ rods, but $\phi=0.07$ for $\alpha=11$. We observed swarming at lower concentrations for higher aspect ratios because of the increased likelihood of longer rods interacting and decreased rotational diffusion. 

As the $\phi$ is increased further, swarms moving in different directions start colliding with each other to form small clusters of rods. Here, we define small clusters as a group of rods with different orientations that lack polar order and are larger than swarms. In this small cluster state, we observe a sudden increase in the maximum cluster size. Hence, we distinguish the swarming and small cluster states using another threshold on maximum cluster size at 25\% of the total number of rods. This threshold is crossed at $\phi=0.32$ for $\alpha=5$ rods but at $\phi=0.22$ for $\alpha=11$ rods (Fig. \ref{fig:simstatediag}). The swarming state not only has swarms (a collection of rods moving in the same direction) but also small clusters (a collection of rods that are not aligned). So, it is noteworthy that the different states of collective dynamics can coexist. 

At even higher area fractions, we observe a state with large clusters, where most rods in the system are in one large cluster. This state is characterized by a second peak in the cluster size distribution at a size comparable to the total number of particles. In this state, number fluctuations are suppressed due to steric hindrance, similar to the large cluster state in the experiments. The existence of such a state with a bimodal cluster size distribution has been theoretically predicted using a mean-field approximation\cite{peruaniNonequilibriumClusteringSelfpropelled2006} and confirmed in simulations \cite{yangSwarmBehaviorSelfpropelled2010}. 

In the simulations, the jammed state is a state in which steric hindrance almost completely suppresses the motion of rods, and the initial state of the simulation does not disappear throughout the simulation. Since the simulation is strictly 2D, the packing limited the maximum $\phi$ we could attain to $\phi\approx0.76$. 

At intermediate area fractions ($\phi \approx 0.28-0.42$), simulations for $\alpha = 7$ reveal a continued increase in clustering but do not show the turbulent state observed in experiments at similar area fractions. In experiments, this state features swirling motion, transient vortices, and negative spatial velocity correlations. The absence of turbulence in our simulations confirms that hydrodynamic interactions, absent from our model, are essential for this state. Previous simulation studies on pusher-type squirmer rods, including hydrodynamic interactions, have also observed turbulent states but at lower aspect ratios\cite{zantopEmergentCollectiveDynamics2022}.

We note that Wensink \emph{et al.}\cite{wensinkMesoscaleTurbulenceLiving2012} report an active turbulent phase in simulations without explicit hydrodynamics, in contrast with our Brownian dynamics simulations.
Two key differences between their model and ours may explain this discrepancy.
First, Wensink \emph{et al.} used a soft, long-range Yukawa potential to model interparticle interactions between rods, which allows particles to slide past each other, avoid jamming, and enable collective motion that mimics fluid-like behavior and leads to turbulence. In contrast, our simulations implement hard-core steric interactions, where rods strictly exclude each other and cannot overlap, resulting in enhanced steric hindrance.
Second, their simulations were performed without Brownian noise and at very high Péclet numbers, focusing on a regime dominated by collisions. The absence of thermal fluctuations can allow coherent, collision-driven vortices to persist and give rise to turbulence.
In contrast, our simulations were designed to closely match the experimental conditions of our system, which operates at a lower Péclet number ($\approx 40$) and is subject to thermal fluctuations.
Thus, the turbulence observed in Ref.\cite{wensinkMesoscaleTurbulenceLiving2012} arises from modeling choices and parameter regimes that differ fundamentally from our experimental system.

Similar to the experimental $\alpha = 4.5$ rods, which do not exhibit a swarming state, the shortest $\alpha=3$ rods in simulations do not exhibit a swarming state. Their high rotational diffusion and weaker shape-induced alignment result in only transient clustering at low area fractions ($\phi \leq 0.25$). In contrast, our experimental $\alpha=4.5$ rods have a much higher rotational diffusion relative to higher aspect ratios (Fig. \ref{fig:exptrotdiff}) compared to the simulations, which results in even shorter-lived transient clusters. 

Overall, the simulated state diagram aligns well with experimental observations. Since we neglect hydrodynamic interactions in our simulations, we do not quantitatively match the experiments with simulations. However, we could still qualitatively reproduce all the features in the experimental state diagram arising from steric interactions and rotational diffusion. 

\subsection{Cluster size distribution in simulations}

The probability distribution of cluster sizes $P(n)$ serves as a quantitative descriptor of collective dynamics in active rods. Figure~\ref{fig:csdlow} shows the normalized cluster size distributions for various concentrations of $\alpha = 7$ rods in the simulation. At low area fractions ($\phi = 0.06$), the distribution is steeply decaying, consistent with an isotropic state dominated by small, short-lived clusters. As the $\phi$ increases to $\phi = 0.17$ and $\phi = 0.35$, a gradual widening of the distribution emerges, indicating the growth of clusters. The threshold at $n/N = 0.04$ is used to mark the boundary between the isotropic and swarming regimes. As the system transitions into the small cluster state, we observe a sudden increase in the maximum cluster size. So, we use a threshold of $n/N = 0.25$ to separate the swarming and small cluster states. At $\phi = 0.46$, the distribution becomes bimodal, signaling the emergence of large clusters containing most of the rods in the system. The appearance of a second peak around $n/N=1$ demarcates the transition to the large-cluster state.

For a higher $\alpha=15$, the $\phi$ at which the isotropic-swarming threshold (of $4\%$) is crossed is lower (Fig. \ref{fig:csd15}). Further, the threshold at $n/N=0.25$ is crossed at $\phi=0.17$ compared to $\phi=0.28$ for the $\alpha=7$ rods. Finally, the bimodal distribution also appears at a lower $\phi=0.36$.

\subsection{Giant number fluctuations in simulations}

Non-equilibrium systems, including wet and dry active matter such as bacteria\cite{zhangCollectiveMotionDensity2010} exhibiting collective behavior, show giant number fluctuations (GNFs), with a scaling \[\frac{\Delta N}{\sqrt{\langle N \rangle}} \propto \langle N \rangle^\beta.\] where \(\beta > 0\). In an isotropic system in equilibrium, the standard deviation of the number of rods is expected to scale as ${\Delta N} \propto \langle N \rangle^{0.5}$, giving \(\beta = 0\).

In our simulations, we vary the subsystem size from \(7.5\sigma\) to \(60\sigma\) and study the scaling of number fluctuations (see Fig. \ref{fig:numfluctsim}). In the isotropic phase, we obtain $\beta=0.03$ since there are minimal interactions in the very dilute system. Increasing the $\phi$ leads to more collective behavior and stronger number fluctuations, with $\beta=0.12$ in the swarming phase at $\phi=0.21$. We obtained a maximum \(\beta =0.243 \pm 0.002\) in the small cluster phase at $\phi=0.35$. Further increasing the concentration to $\phi= 0.53$ causes a transition to the large cluster state. In this state, most rods are present in one large cluster, and the steric hindrance leads to a reduction in number fluctuations, even though $\beta$ is similar. Finally, fluctuations are again completely suppressed in the jammed state ($\phi=0.71$). 

Our findings are consistent with previous simulation studies on self-propelled rods that have reported exponents corresponding to \(\beta \approx 0.3\) \cite{ginelliLargeScaleCollectiveProperties2010}. In comparison with our experiments, which show maximum GNFs in the turbulent phase at $\phi=0.39$, we still obtain maximum GNFs around $\phi=0.35$ despite not having turbulence. This is because lower area fractions have a lesser number of particles interacting with each other, while higher area fractions have stronger steric hindrance, preventing number fluctuations.

\subsection{Velocity autocorrelation function in simulations}

In our simulations of self-propelled rods with $\alpha=7$, the velocity autocorrelation function (VACF) quantifies how rapidly the system loses its velocity memory. In the isotropic phase, rods lose the correlation in their velocity mainly because of rotational diffusion. As we increase the $\phi$ and enter the swarming phase, the VACF exhibits a faster decay with increasing $\phi$ (see Fig. \ref{fig:vacfsim}), primarily due to an enhanced frequency of steric collisions. In our experimental system, hydrodynamic interactions mediated by the surrounding fluid further accelerate the decay in VACF, leading to a more pronounced reduction in velocity correlations. Even though there are no hydrodynamic interactions in the simulations, we can capture the underlying mechanism by which increased particle density disrupts the persistence of individual rod motion.

{
\subsection{Finite element method simulations} \label{comsol}

The flow generated by catalytically active rods arises from complex physicochemical interactions between solutes, surfaces, and the surrounding fluid (see Fig. \ref{fig:velocity_vs_distance_silica_1um_errorbar},\ref{fig:Flowfield_AR_7.5_bare_coated_1um}). Because the detailed mechanisms are difficult to capture explicitly \cite{mu2022light}, we employ a phenomenological model to reproduce the experimentally observed flow fields around a stationary rod near a planar wall. The model is based on the framework of neutral diffusiophoresis \cite{katuriInferringNonequilibriumInteractions2021} and is implemented using finite-element simulations in \textsc{COMSOL Multiphysics 6.3}.

A three-dimensional geometry is constructed to represent an immobilized catalytic rod near a planar substrate. The rod consists of a catalytic head, modeled as an ellipsoid with semi-axis lengths $(0.25,\,0.5,\,0.5)\,\mu\mathrm{m}$, and an inert body, represented by a coaxial cylinder of radius $0.35\,\mu\mathrm{m}$ and length $5\,\mu\mathrm{m}$. The rod’s long axis lies parallel to the substrate, and the bottom surface of the rod is at a height of $0.2\,\mu\mathrm{m}$ from the substrate. The surrounding fluid domain is a right circular cylinder of radius $10\,\mu\mathrm{m}$ and height $10\,\mu\mathrm{m}$, centered on the particle.

The catalytic head emits a neutral solute at a uniform rate $\kappa$. The solute diffuses in the surrounding medium with diffusion coefficient $D$, establishing a concentration field $c(\mathbf{x})$ governed by the transport equation,
\begin{equation}
\nabla \cdot \mathbf{J} + \mathbf{u} \cdot \nabla c = 0, 
\qquad 
\mathbf{J} = -D \nabla c.
\label{eq:diffusion}
\end{equation}
Boundary conditions for the concentration field are imposed as follows:
\begin{align}
- D\,\mathbf{n}\cdot\nabla c &= \kappa && \text{on the catalytic (active) head}, \\
\mathbf{n}\cdot(\mathbf{J}+ \mathbf{u} c) &= 0 && \text{on inert surfaces (rod and wall)}, \\
c &\rightarrow 0 && \text{at the outer (far-field) boundaries},
\end{align}
where \(\mathbf{n}\) is the outward unit normal.
Gradients in the solute concentration generate tangential slip velocities on the solid surfaces, described by
\begin{equation}
\mathbf{u}_\text{s}(\mathbf{s}) = - b(\mathbf{s}) \nabla_{\parallel} c(\mathbf{s}),
\label{eq:slip}
\end{equation}
where $\nabla_{\parallel}$ denotes the tangential surface gradient and $b(\mathbf{s})$ is the local surface mobility, a material-dependent parameter encapsulating solute–surface interaction effects:
\[
b(\mathbf{s}) =
\begin{cases}
b_\text{c}, & \text{on the catalytic surface (head)},\\
b_\text{i}, & \text{on the inert surface (rod)},\\
b_\text{w} & \text{on the wall}.
\end{cases}
\]
The surface flows drive bulk fluid motion governed by the incompressible Stokes equations:
\begin{equation}
- \nabla p + \mu \nabla^2 \mathbf{u} = 0, 
\qquad 
\nabla\cdot\mathbf{u} = 0,
\label{eq:stokes}
\end{equation}
where $\mathbf{u}$ and $p$ denote the velocity and pressure fields, respectively, and $\mu$ is the dynamic viscosity of the fluid. Here, all external boundaries, except the planar substrate, are treated as open boundaries with zero normal stress.

At this point, the flow field $\mathbf{u}$ consists of hydrodynamic flow driven by both phoretic slip on the particle surface, $\mathbf{u_\text{rod}}$, and chemiosmotic slip on the planar bottom wall, $\mathbf{u_\text{wall}}$. To compare simulated flow fields with experimentally observed tracer motion, the phoretic drift of tracer particles is included as
\begin{equation}
\mathbf{u}_{\text{total}} = \mathbf{u} + \mathbf{u}_{\text{phoretic}} = \mathbf{u_\text{rod}} + \mathbf{u_\text{wall}} + b_{\text{tracer}} \nabla c,
\end{equation}
where $b_{\text{tracer}}$ is the tracer mobility. The total tracer velocity $\mathbf{u}_\text{total}$ thus includes both the hydrodynamic and osmotic flow, as well as the phoretic contribution. The COMSOL model parameters are summarized in Table~\ref{tab:params}.

The COMSOL model, based on the framework of neutral diffusiophoresis, successfully captures near-wall effects and reproduces the qualitative features of the experimentally observed flow fields (fig. \ref{fig:Flowfield_AR_7.5_and_15.1_silica_tpm}), without explicitly resolving electrokinetic or other detailed physicochemical interactions. When the phoretic motion of the tracer particles is included, the simulated flow field in the XY plane (fig.~\ref{fig:comsol}A) resembles the experimentally measured flow field parallel to the wall, with the exception of the phoretic attraction observed experimentally behind the catalytic head (fig. \ref{fig:Flowfield_AR_7.5_and_15.1_silica_tpm}). This discrepancy suggests that the experimental system involves additional physicochemical effects not represented in the simplified neutral diffusiophoretic model. We consider the XY plane through the center of the rod, at a height of $0.55\, \mu m$ from the substrate, comparable to the size of tracers.

In order to better understand the effect of the wall, we look at osmotic flows in the XZ plane, perpendicular to the wall (fig.~\ref{fig:comsol}B). The simulation yields inflow from above the rod and outflow near its front, qualitatively replicating the experimental osmotic slip flows induced along the wall (see movie S2).

\newpage

\begin{figure}[!hbt]
    \centering
    \includegraphics[width=0.99\linewidth]{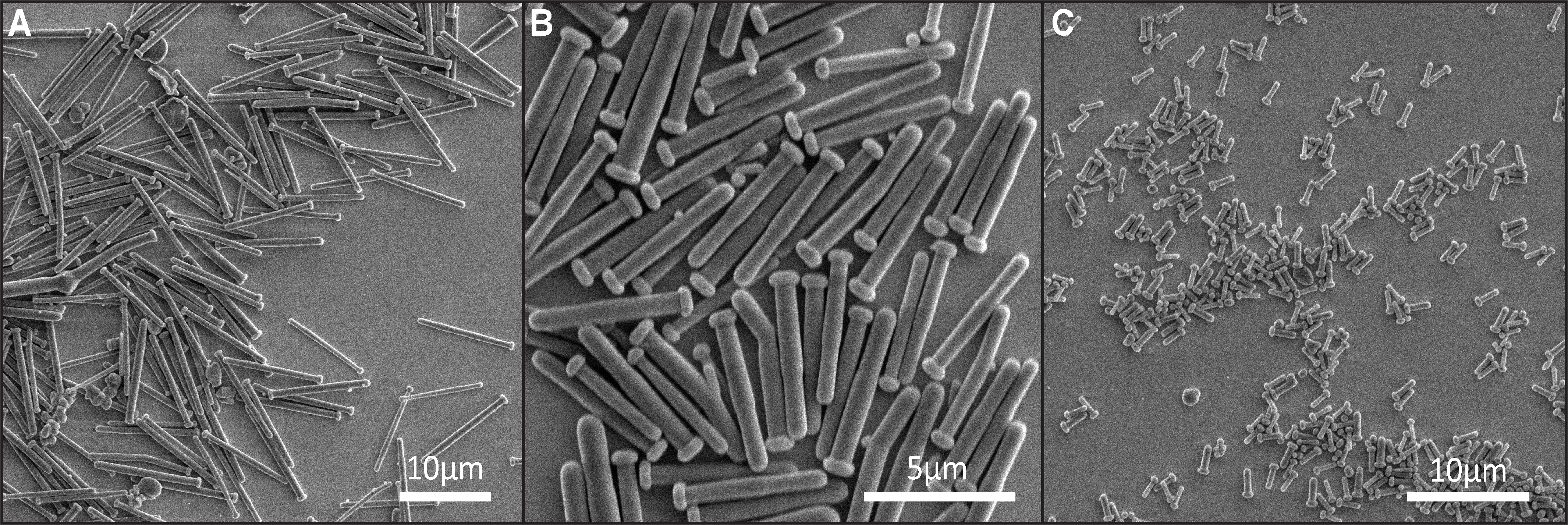}
    \caption{\textbf{Scanning electron microscope images of different aspect ratios rods.} (A) $\alpha = 4.5\pm 0.34$ ($l = 2 \pm 0.1$ $\mu\mathrm{m}$ and $d = 0.45 \pm 0.05$ $\mu\mathrm{m}$).
  (B) $\alpha = 7.5 \pm 1.8$ ($l = 5.35 \pm 0.55$ $\mu\mathrm{m}$ and $d = 0.71 \pm 0.14$ $\mu\mathrm{m}$).
  (C) $\alpha = 15.1 \pm 2.15$ ($l = 10.9 \pm 0.72$ $\mu\mathrm{m}$ and $d = 0.72 \pm 0.17$ $\mu\mathrm{m}$).}
    \label{fig:SEM}
\end{figure}

\begin{figure}[htbp!]
	\centering
	\includegraphics [width=0.65\textwidth]{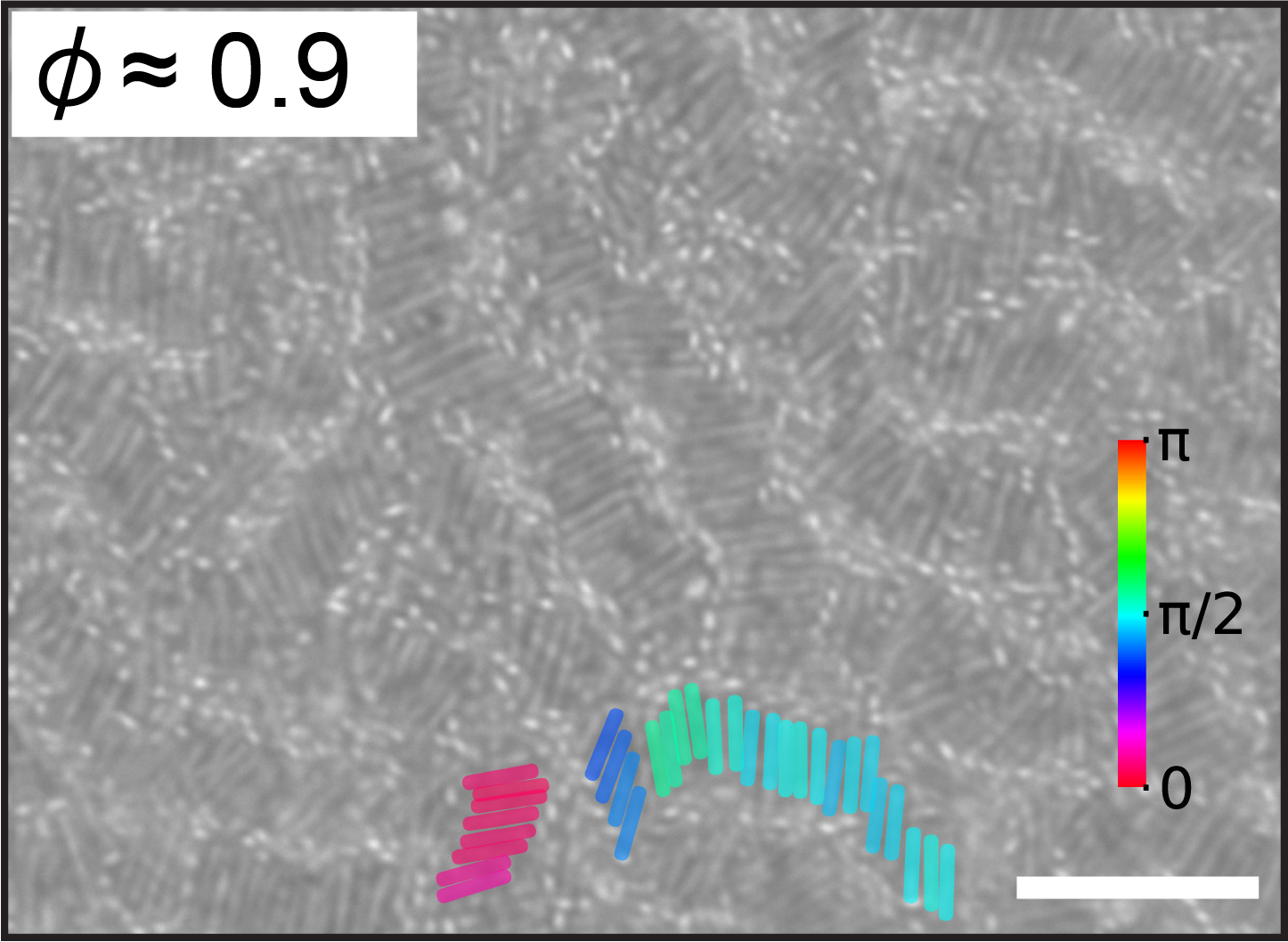} 
	\caption{\YS{\textbf{Jammed state of active rods with aspect ratio $\alpha = 7.5$ at $\phi \approx 0.9$.} Overlaid rod orientations (in radians) illustrate local nematic ordering within the densely packed jammed configuration. Scale bar: 10~$\mu$m.}}
    
	\label{fig:1}
\end{figure}

\begin{figure}[H]
    \centering
    \includegraphics[width=0.7\linewidth]{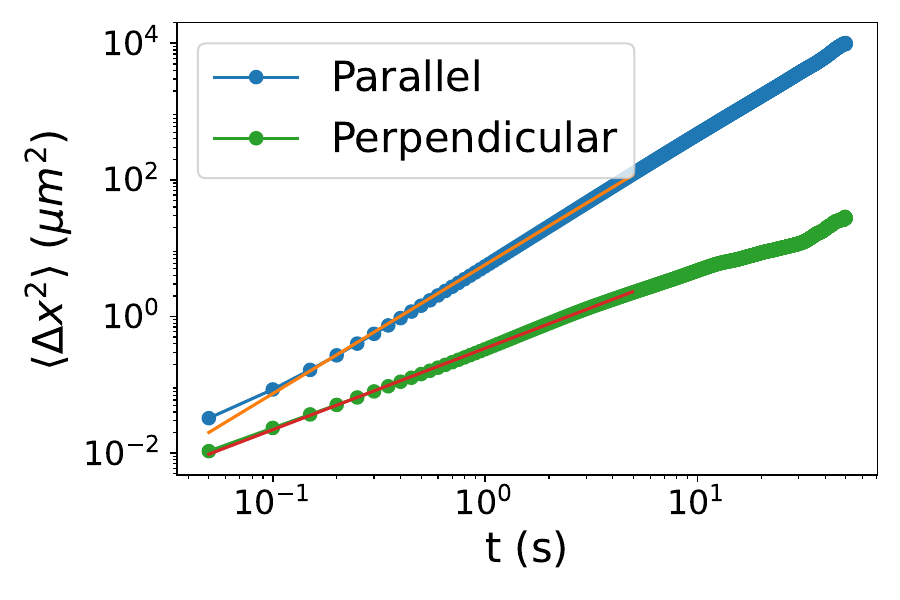}
    \caption{\textbf{Mean square displacement (MSD) of rods with  $\alpha = 7.5$, calculated in the directions parallel and perpendicular to the rod long axis from experimental trajectories of rods.} The fitted lines to the MSD curves show slopes of 1.88 (parallel) and 1.19 (perpendicular), indicating anisotropic translational diffusion. The nearly ballistic scaling in the parallel direction reflects persistent motion along the rod long axis.}
    
    \label{fig:exptmsdparperp}
\end{figure}

\begin{figure}[H]
    \centering
    \includegraphics[width=0.65\linewidth]{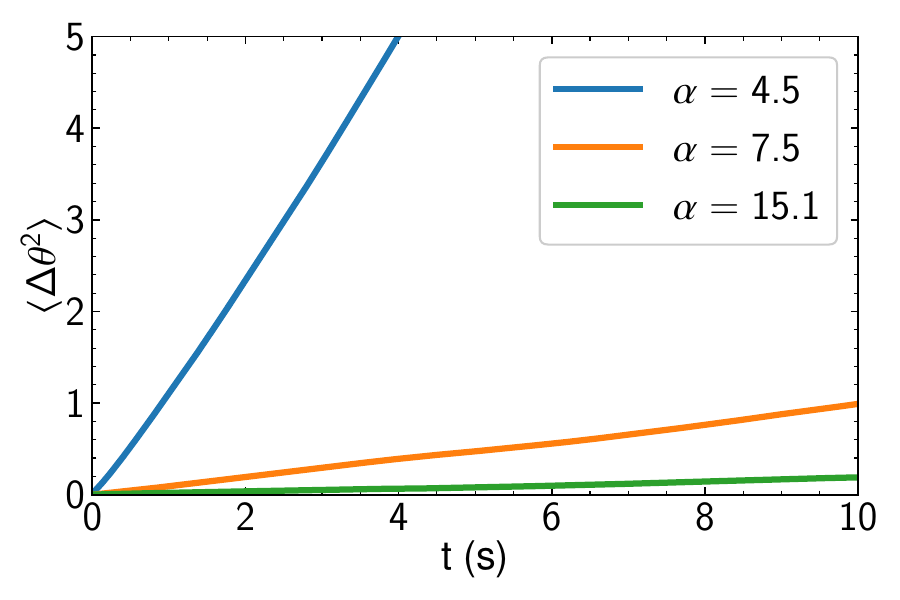}
    \caption{\textbf{Mean square angular displacement $\langle \Delta \theta^2 \rangle$ as a function of time $t$ for rods with aspect ratios $\alpha = 4.5$, $7.5$ and $15.1$ from experiments.} The slope of each curve indicates the rotational diffusion coefficient $D_R$, which decreases with increasing $\alpha$. Rods with $\alpha = 4.5$ exhibit the highest rotational diffusion with $D_R = 0.662$ $\mathrm{rad}^2/\mathrm{s}$. For $\alpha = 7.5$, the diffusion slows significantly with $D_R = 0.048$ $\mathrm{rad}^2/\mathrm{s}$, and further decreases for $\alpha = 15.1$ to $D_R = 0.009$ $\mathrm{rad}^2/\mathrm{s}$ showing the strong dependence of $D_R$ on rods shape anisotropy.}
    \label{fig:exptrotdiff}
\end{figure}

\begin{figure}[H]
    \centering
    \includegraphics[width=0.5\linewidth]{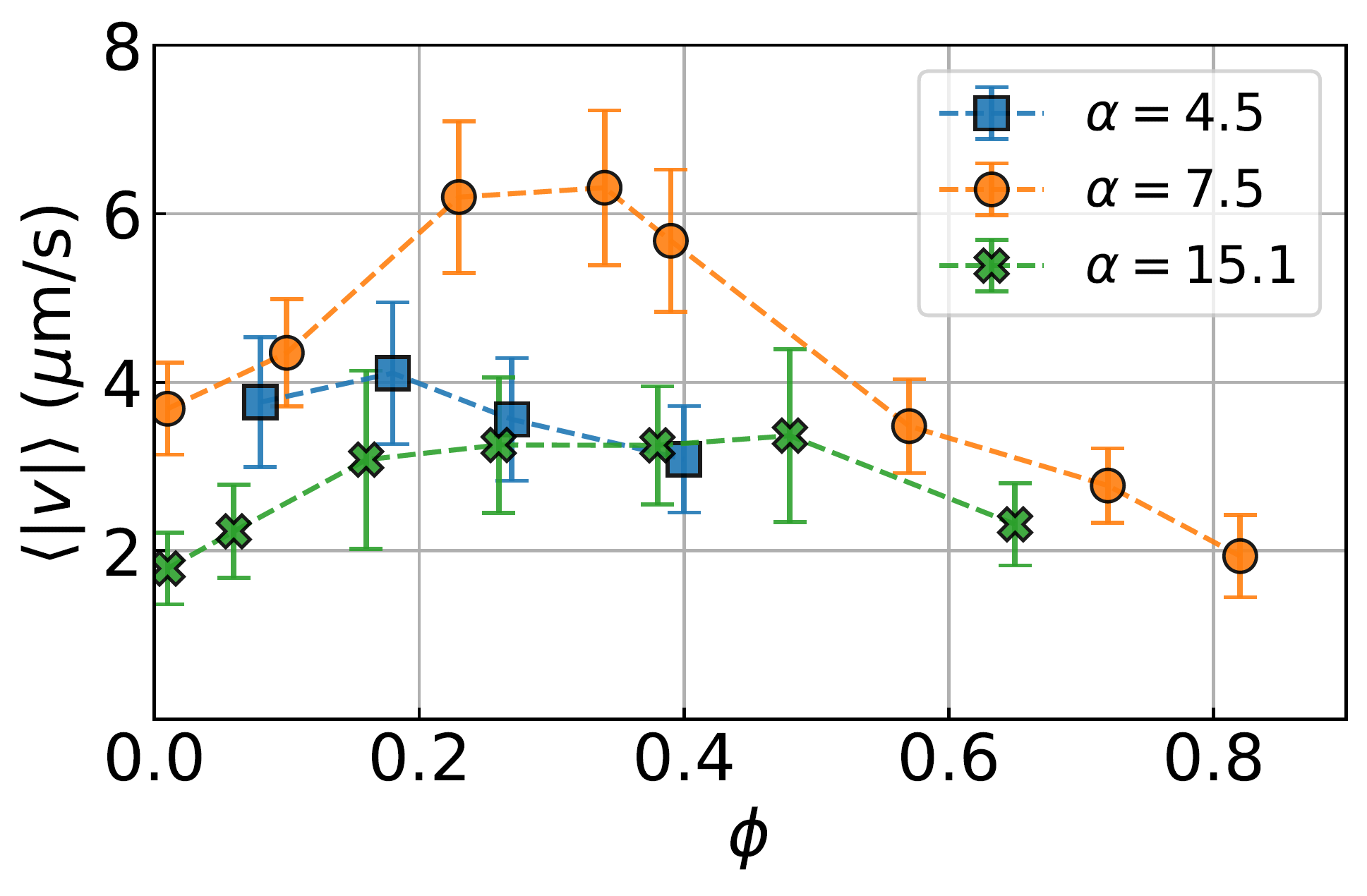}
    \caption{\textbf{Experimental mean speed of active rods for various $\alpha$ with $\phi$.} A non-monotonic trend is observed for all aspect ratios: at low $\phi$, velocities remain low because rod-rod interactions are minimal. As $\phi$ increases, the mean speed also increases at intermediate concentrations, most likely caused by hydrodynamic interactions and enhanced fuel mixing. However, at higher $\phi$, crowding and frequent collisions suppress rod movement, causing the mean speed to decline again. The error bars show the standard deviation of the mean speed.}
    \label{fig:Velocity}
\end{figure}

\begin{figure}[H]
    \centering
    \includegraphics[width=\linewidth]{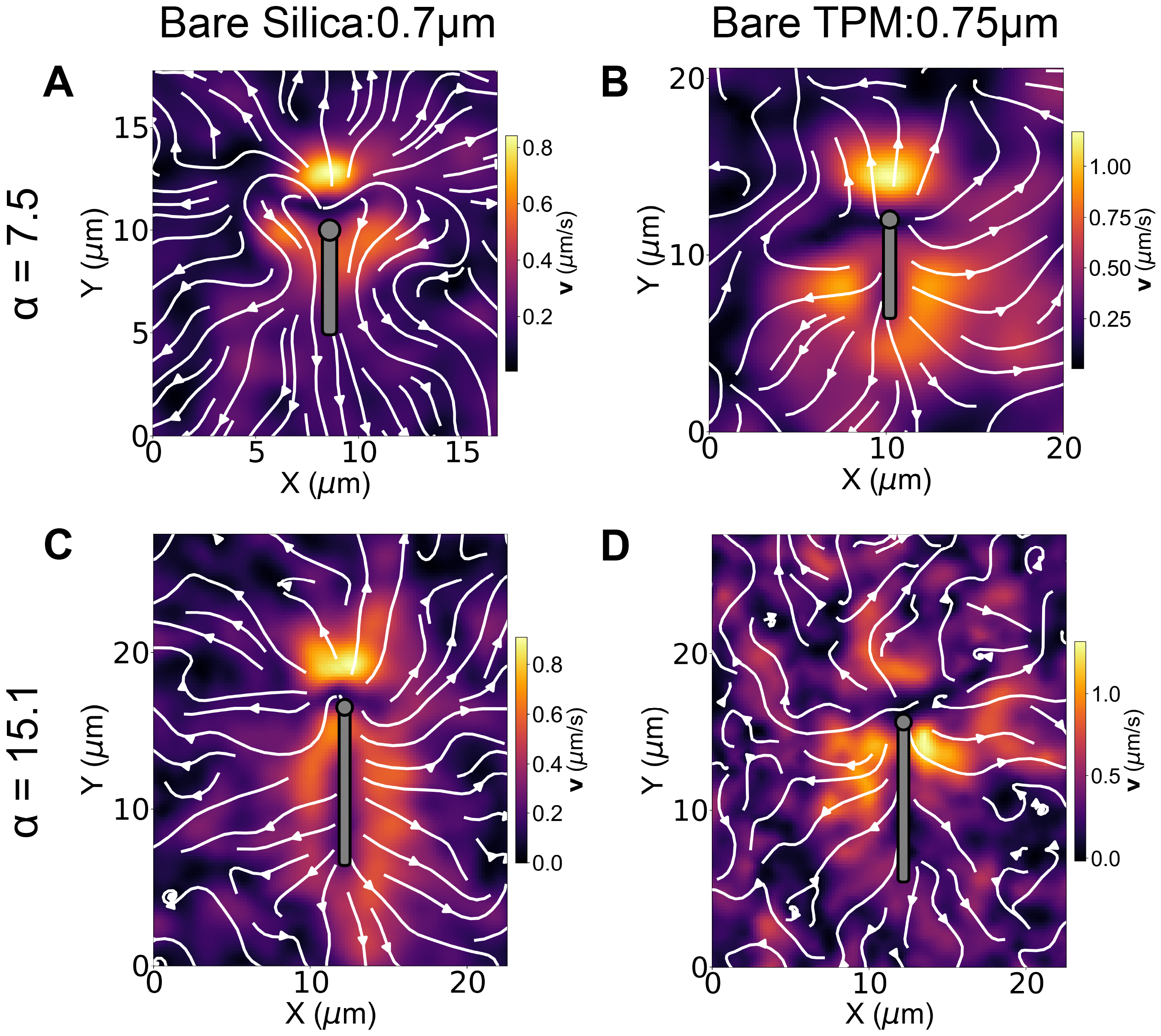}
    \caption{\YS{\textbf{Flow streamlines and their magnitude around active rods with different aspect ratios and tracer types.} (A) $\alpha = 7.5$; flow field obtained with 0.7 $\mu$m bare silica tracers.  (B) $\alpha = 7.5$; flow field obtained with  0.75~$\mu$m bare TPM tracers. (C) $\alpha = 15.1$; flow field obtained with 0.7 $\mu$m bare silica tracers.  (D) $\alpha = 15.1$; flow field obtained with 0.75~$\mu$m bare TPM tracers.}}
    \label{fig:Flowfield_AR_7.5_and_15.1_silica_tpm}
\end{figure}

\begin{figure}[H]
    \centering
    \includegraphics[width=0.6\linewidth]{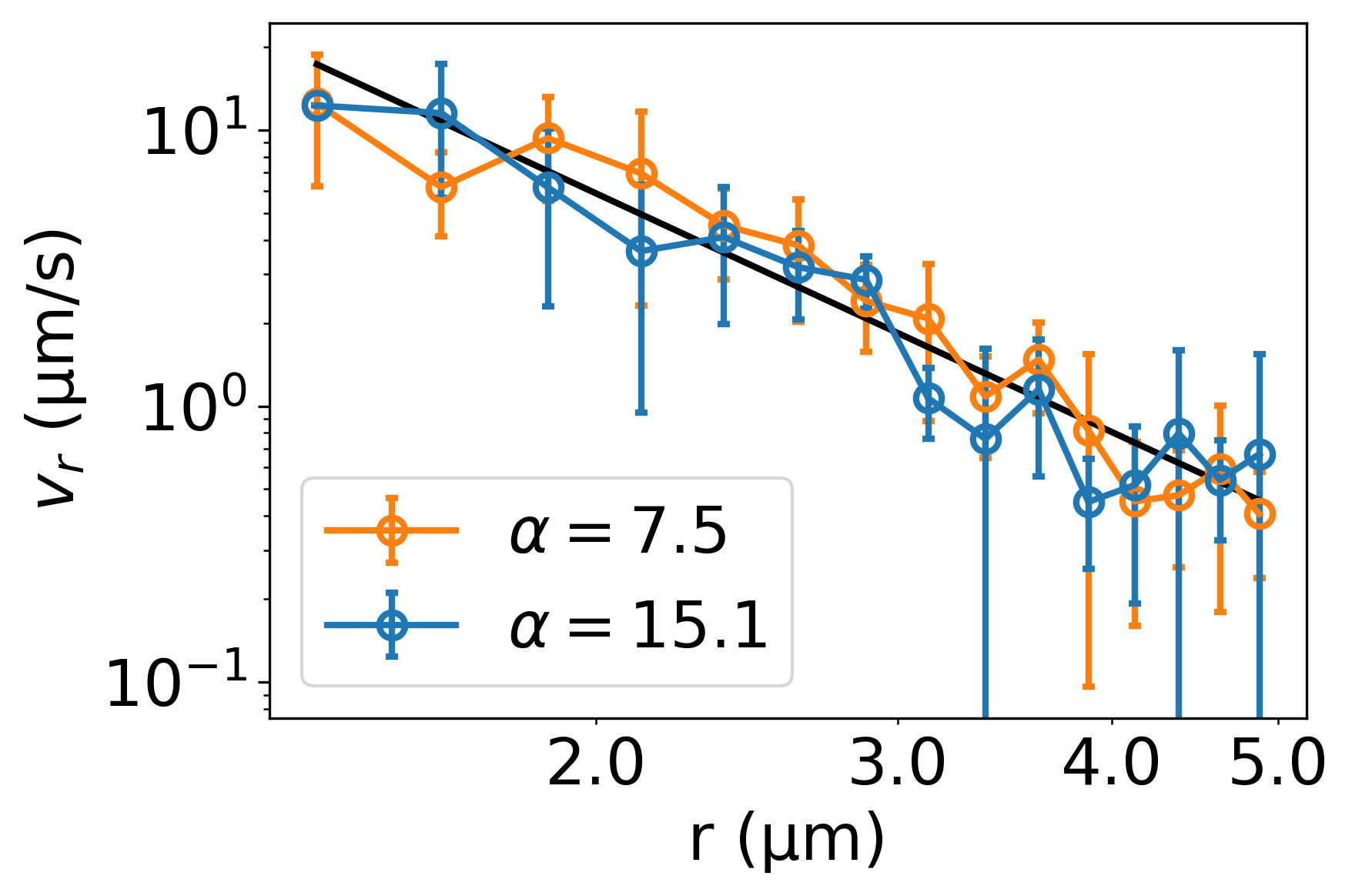}
    \caption{\YS{\textbf{The near-field flow generated by the active rods remains unchanged with aspect ratio.} Bare silica 0.7~$\mu$m tracer velocity ($v_r$) versus distance $r$ from the center of the reactive head of immobilized rods with different aspect ratios: $\alpha = 7.5$ (orange) and $\alpha = 15.1$ (blue). Both datasets collapse onto a similar near-field scaling, where $v(r)$ decays as $1/r^3$, primarily determined by the identical head size where the catalytic reaction occurs. A bin size of 0.25~$\mu$m was used to bin instantaneous tracer velocities. The black line is a fit with an exponent of -2.87.}}
    \label{fig:velocity_vs_distance_silica_700nm_errorbar}
\end{figure}

\begin{figure}[H]
    \centering
    \includegraphics[width=0.6\linewidth]{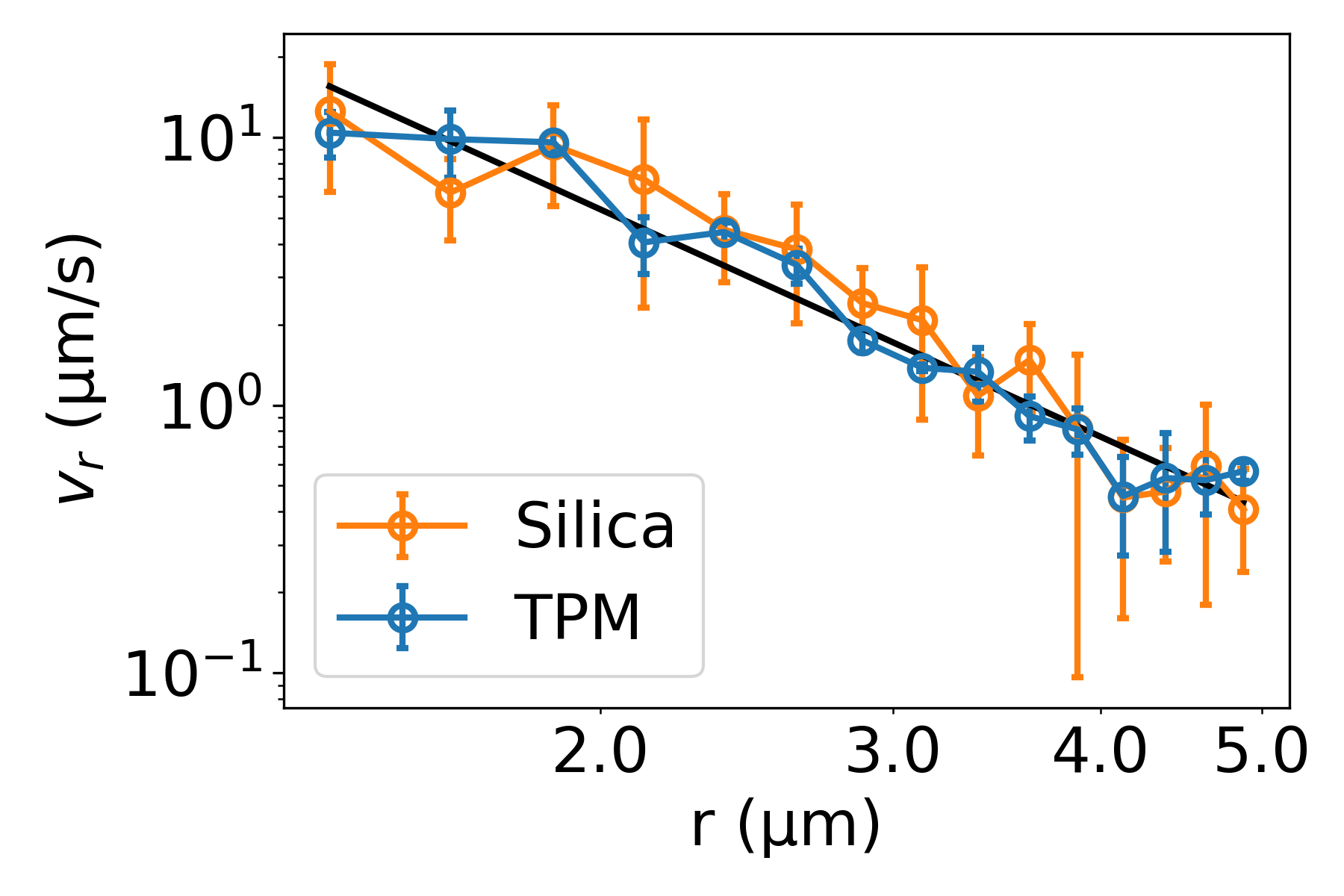}
    \caption{\YS{\textbf{Decay of radial component of tracer velocities, $v_r$ vs distance $r$ in front of rod head for 0.75~$\mu$m TPM tracers (blue) and 0.7~$\mu$m silica tracers (orange).} A bin size of 0.25~$\mu$m was used to bin instantaneous tracer velocities. Error bars show standard deviation across multiple rods. The black line represents a fit with an exponent of -2.82.}}
    \label{fig:velocity_vs_distance_tpm}
\end{figure}

\begin{figure}[H]
    \centering
    \includegraphics[width=\linewidth]{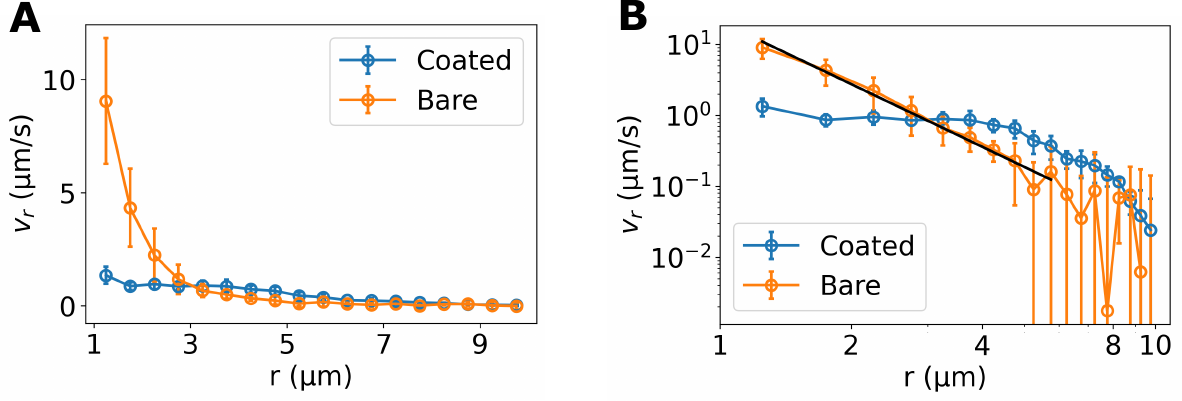}
    \caption{\YS{\textbf{Decay of radial component of tracer velocities, $v_r$ vs distance $r$ in front of rod head for bare and pNIPAM coated 1.04~$\mu$m silica tracers.} (A) Bare tracer (orange) velocities are stronger in the near-field dominated by phoretic interactions. In contrast, brush-coated sterically stabilized tracer (blue) suppresses phoresis, resulting in much lower near-field velocities and weaker osmotic flows. A bin size of 0.5~$\mu$m was used to bin instantaneous tracer velocities. Error bars show standard deviation across multiple rods. (B) Log-log plot of $v_r$ vs r showing a scaling close to $1/r^3$ in the case of bare tracers. The black line shows a fit with exponent -2.93.}}
    \label{fig:velocity_vs_distance_silica_1um_errorbar}
\end{figure}

\begin{figure}[H]
    \centering
    \includegraphics[width=\linewidth]{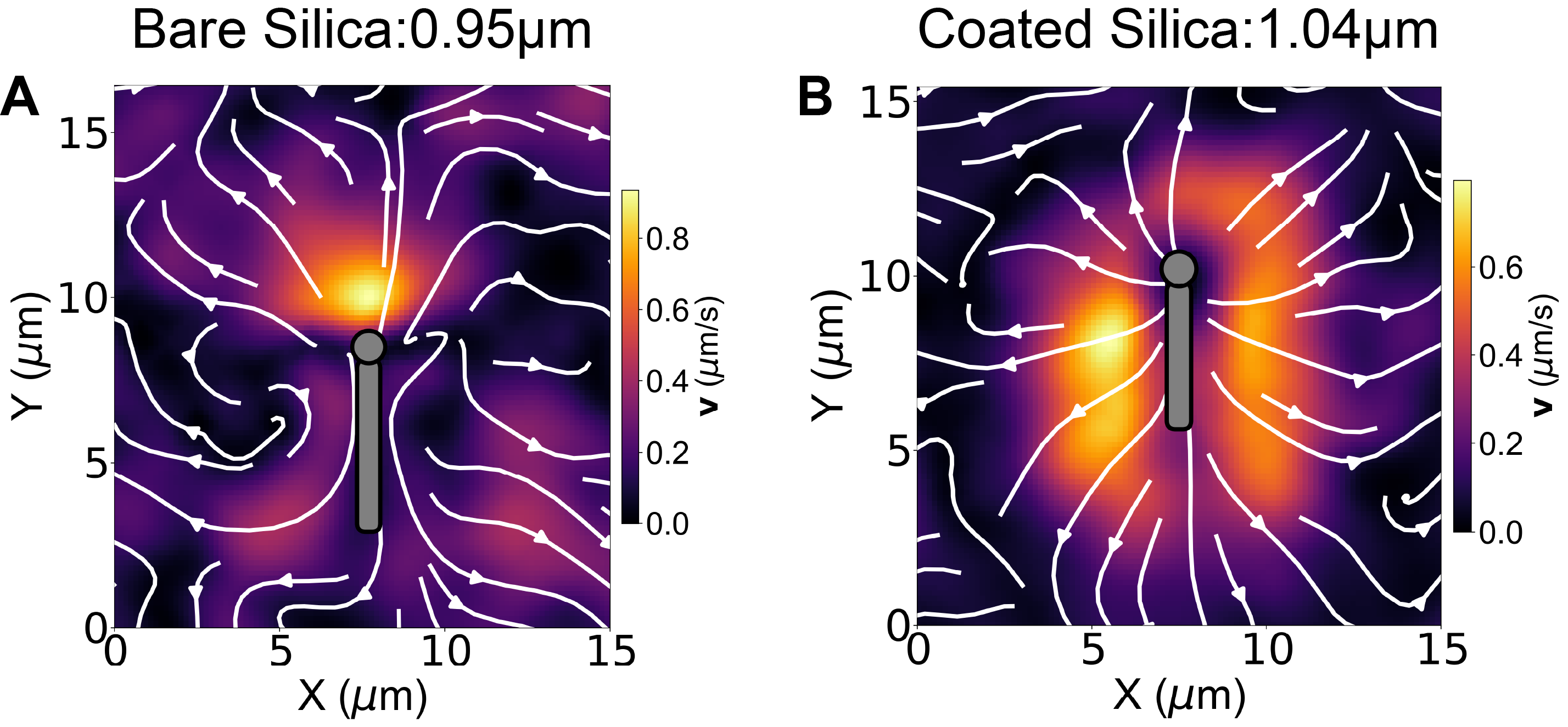}
    \caption{\YS{\textbf{Flow streamlines and their magnitude around $\alpha = 7.5$ active rods with bare (uncoated) and coated tracers.} (A) flow field obtained with 0.95~$\mu$m bare silica tracers. (B) flow field obtained with 1.04~$\mu$m coated silica tracers.}}
    \label{fig:Flowfield_AR_7.5_bare_coated_1um}
\end{figure}

\begin{figure}[H]
    \centering
    \includegraphics[width=0.6\linewidth]{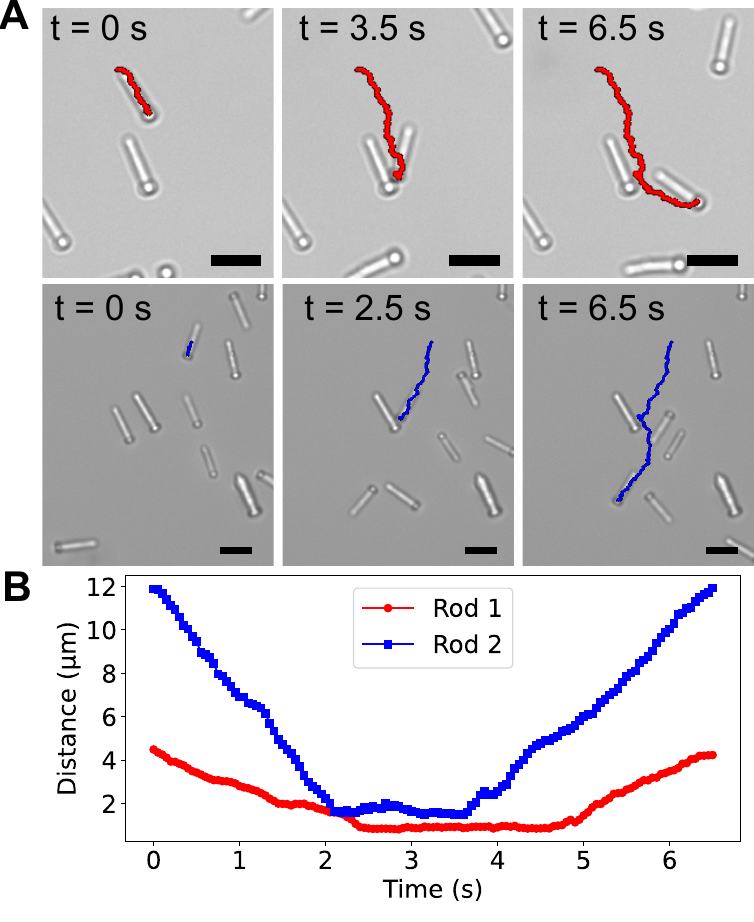}
    \caption{\YS{\textbf{Interactions between immobilized and freely moving rods.} (A) Time-lapse snapshots from two separate experiments demonstrate interactions between immobilized and freely moving rods, confirming weak attraction at the rear of the head. (B) Corresponding plot displays the pair distance between the moving rod and the immobilized rod over time as it approaches, remains briefly behind the head for $\sim$2 s, and then departs. Scale bars represent 5~$\mu$m.}}
    \label{fig:rod-rod}
\end{figure}

\begin{figure}[H]
\centering
\includegraphics[width=\linewidth]{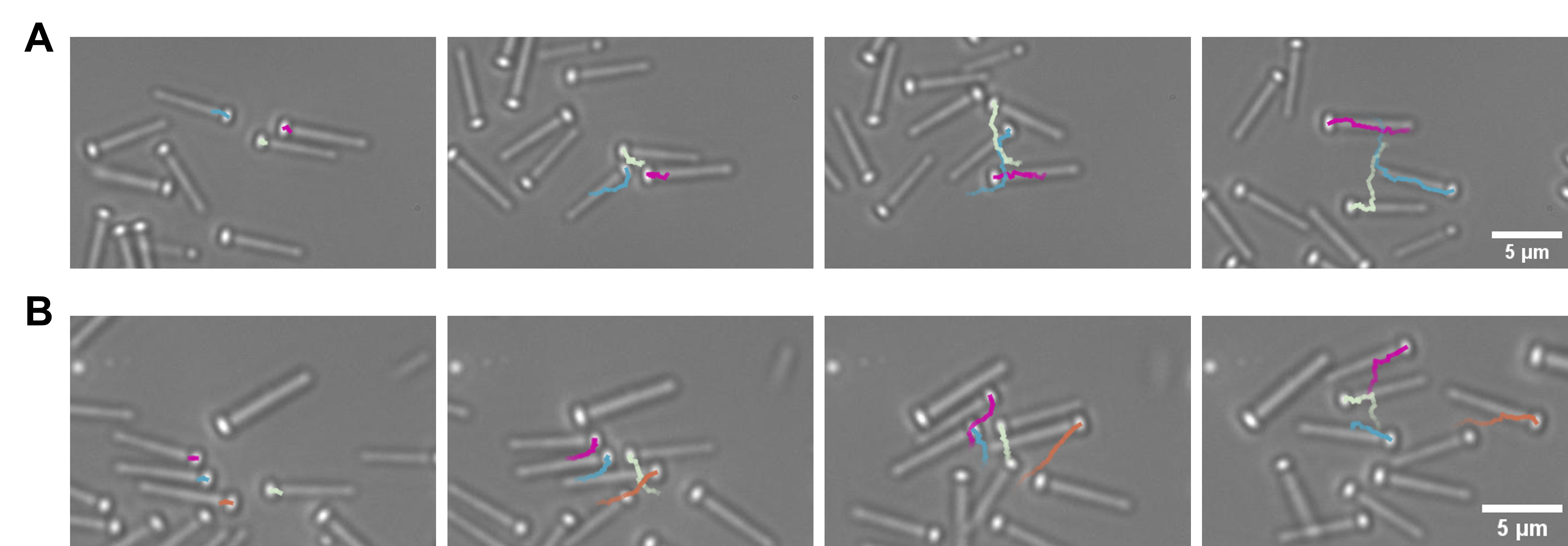}
\caption{\YS{\textbf{Interactions between freely moving active rods.} Time-lapse snapshots from two separate events showing head-on interactions between freely moving active rods, confirming that phoretic attractions are weak. Scale bars represent 5~$\mu$m.}}
\label{fig:rod-rod-free}
\end{figure}

\begin{figure}[H]
    \centering
    \includegraphics[width=0.8\linewidth]{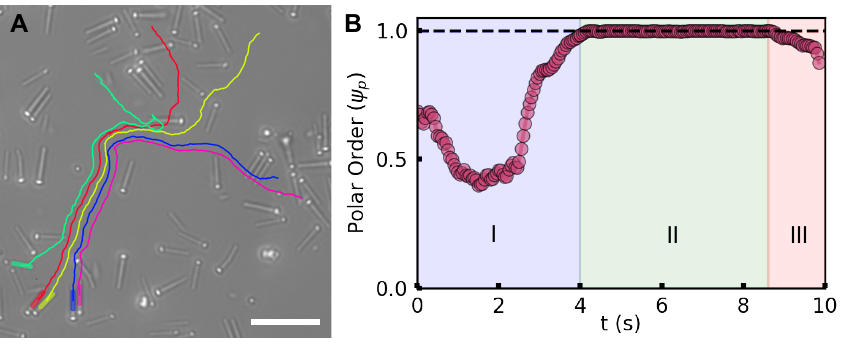}
    \caption{\textbf{Evolution of a single swarm.} (A) Microscope image with overlaid rod trajectories for 10 seconds of $\alpha=7.5$ rods at $\phi=0.11$ illustrates how rods from different orientations come together, synchronize their orientation into a small cluster, and then separate. (B) Local polar order parameter ($\psi_{p}$) for the rods tracked in (A), showing orientation alignment progression with time from disordered motion with low $\psi_{p}$ (region I, $t=0$--2\,s) to a highly aligned state with $\psi_{p}\approx 1$ (region II, $t=2$--8.5\,s), and finally cluster splitting characterized by a drop in $\psi_{p}$ (region III, $t=8.5$--10\,s). Scale bar is 10 $\mu$m.}
    \label{fig:Single_swarm}
\end{figure}

\begin{figure}[H]
    \centering
    \includegraphics[width=\linewidth]{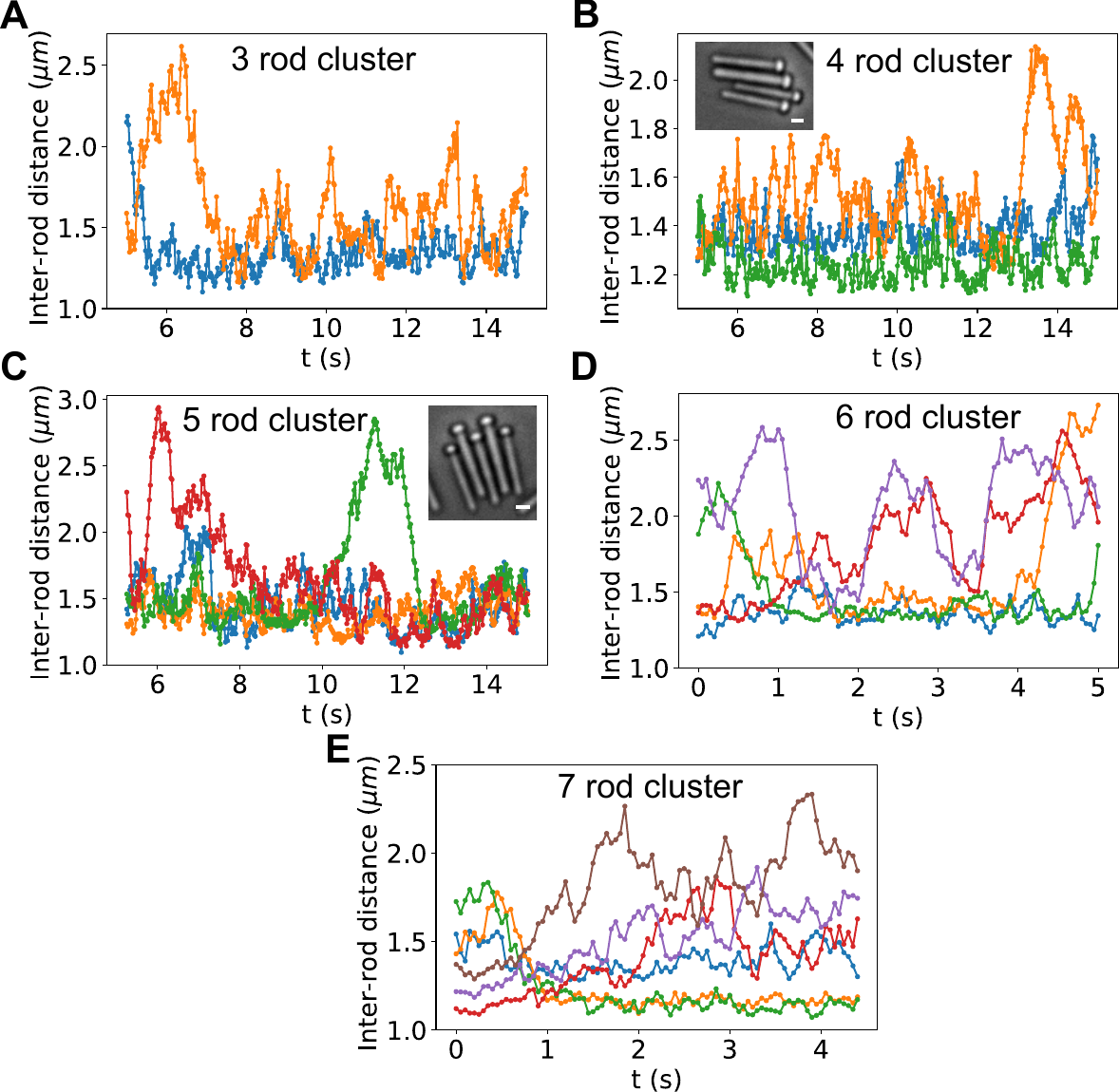}
    \caption{\YS{\textbf{Fluctuations in the distance between heads of nearest neighbor pairs in clusters for different cluster sizes.} Snapshots of representative clusters, with a scale bar indicating 1~$\mu$m, are presented in the insets.}}
    \label{fig:fluct_comb}
\end{figure}

\begin{figure}[H]
    \centering
    \includegraphics[width=\linewidth]{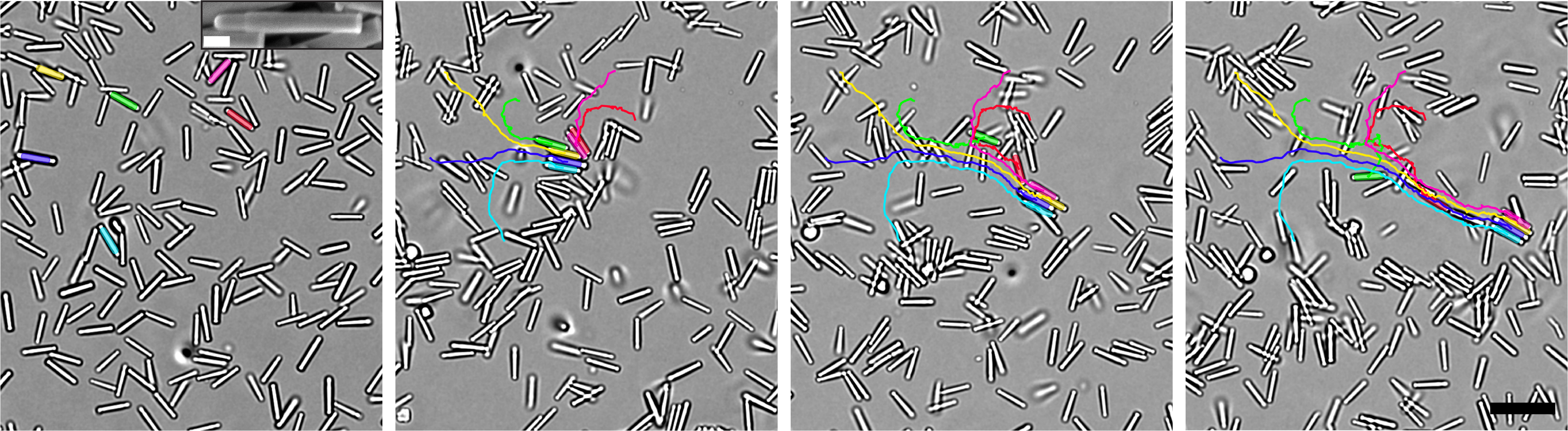}
    \caption{\YS{\textbf{Collective motion of rods with equal head-tail size.} Time-lapse (0–60~s) combined fluorescence and bright-field microscopy snapshots of rods with head size comparable to the tail, and aspect ratio $\alpha = 7.5$, at area fraction $\phi = 0.12$. (Inset: SEM image of a rod with equal head and tail. Scale bar: 1~$\mu$m.) The sequence illustrates the temporal evolution of the rod swarm, with overlaid trajectories highlighting the swarming motion. Snapshots are contrast-enhanced for clarity. Scale bar: 10~$\mu$m.}}
    \label{fig:Small_head_rods}
\end{figure}
\begin{figure}[H]
    \centering
    \includegraphics[width=\linewidth]{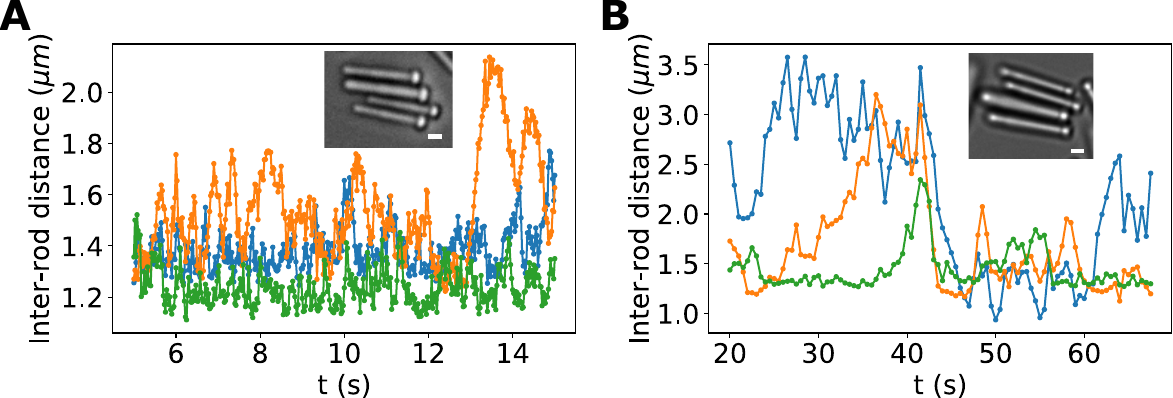}
    \caption{\YS{\textbf{Fluctuations in the distance between heads of nearest neighbor pairs in clusters.} Insets show representative cluster snapshots with a scale bar of 1 $\mu$m. (A) for $\alpha = 7.5$ with head size~$0.9\,\mu\mathrm{m}$ and  tail size $0.7~\mu$m. (B) for $\alpha = 7.5$ with equal head and tail size of 0.73~$\mu$m. }}
    \label{fig:fluct_smooth}
\end{figure}

\begin{figure}[H]
    \centering
    \includegraphics[width=\linewidth]{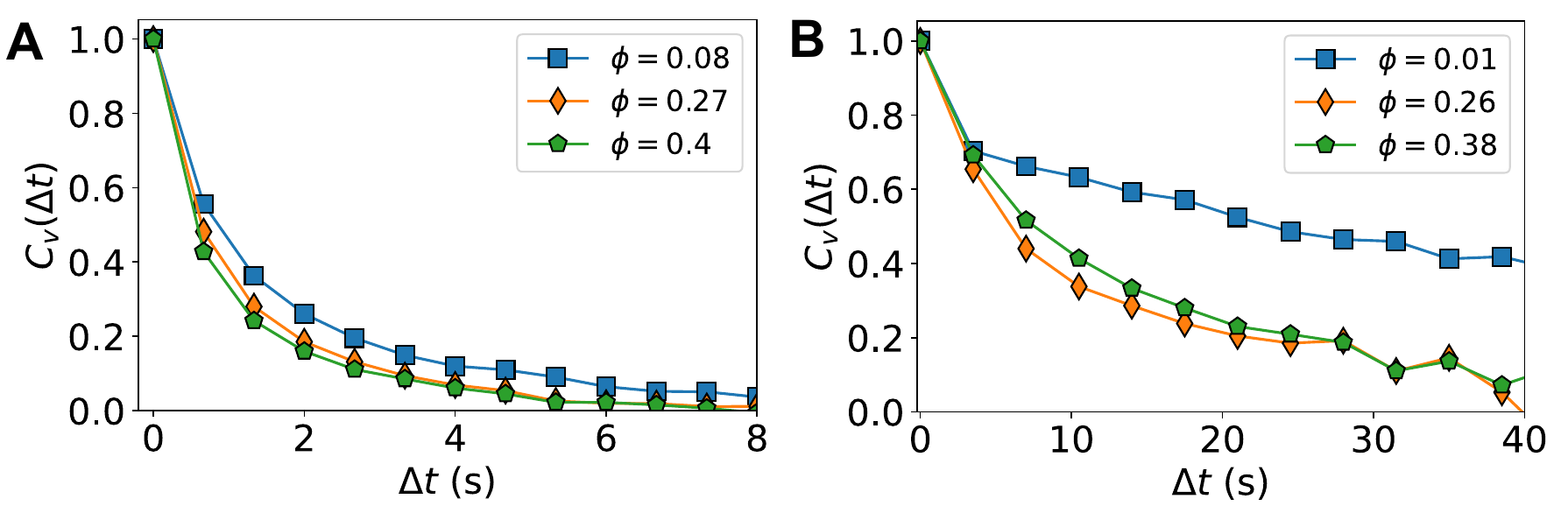}
    \caption{\textbf{Velocity autocorrelation function (VACF) from experiments of active rods with (A) aspect ratio \( \alpha = 4.5 \) and (B) \( \alpha = 15.1 \).} In both cases, the isotropic phase exhibits the slowest decay, where loss of velocity correlation is primarily governed by rotational diffusion. The slower decay in (B) compared to (A) reflects the lower rotational diffusion \(D_r\) associated with the higher aspect ratio. As concentration increases, the VACF decays more rapidly in both cases due to enhanced inter-rod interactions. However, for \( \alpha = 4.5 \), rods form only short-lived transient clusters, and the decay remains relatively fast. In contrast, \( \alpha = 15.1 \) rods exhibit swarming and flocking behavior, which result in similar VACF decay profiles across these ordered states.}
    \label{fig:exptvacf}
\end{figure}

\begin{figure}[H]
    \centering
    \includegraphics[width=0.99\linewidth]{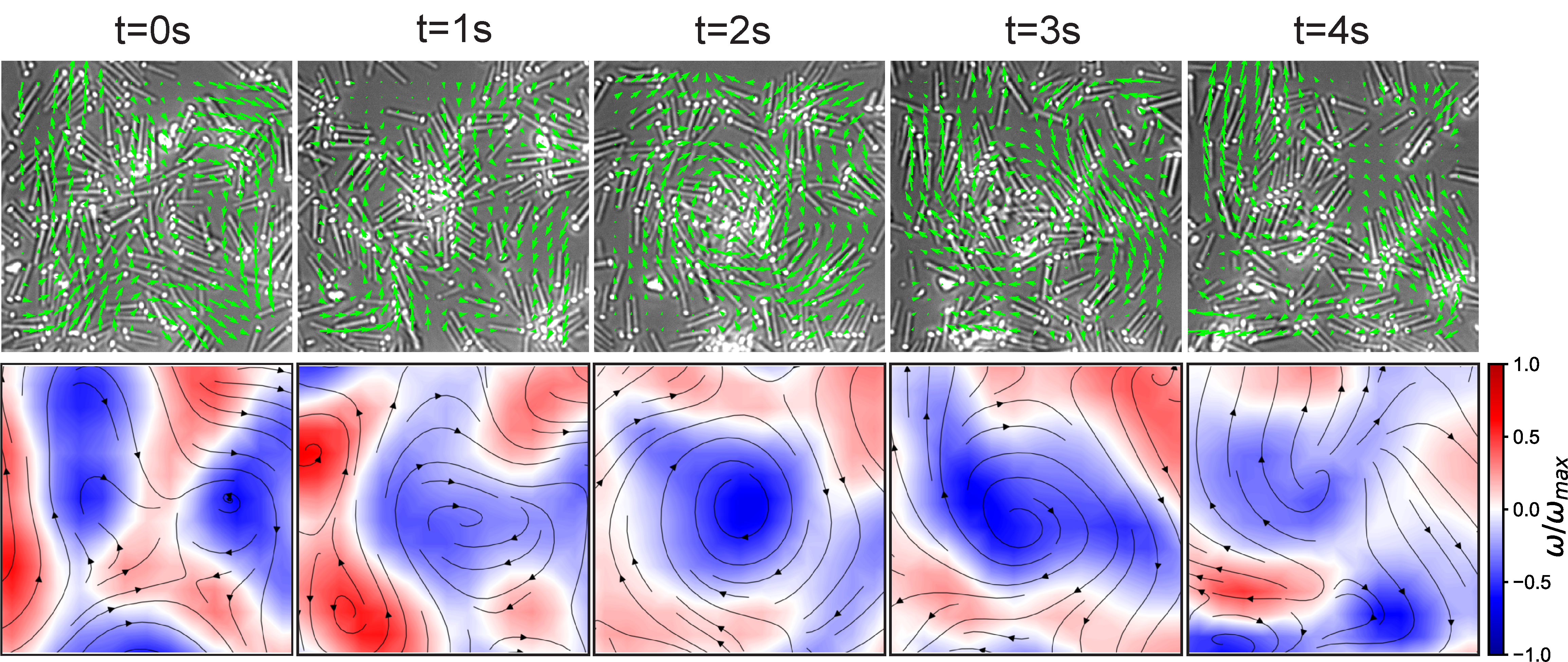}
    \caption{\textbf{Evolution of a vortex in the turbulent phase.} Top panel: Instantaneous velocity vectors for $\alpha=7.5,\phi=0.39$; bottom panel: Corresponding vorticity field. Together, they reveal that rod clusters start to converge at t = 0 s and collide with each other at t = 1 s; upon colliding, they develop a vortex at t = 2 s, which eventually dissolves at t = 4 s.}
    \label{fig:Single_vortex}
\end{figure}

\begin{figure}[!hbt]
    \centering
    \includegraphics[width=0.7\linewidth]{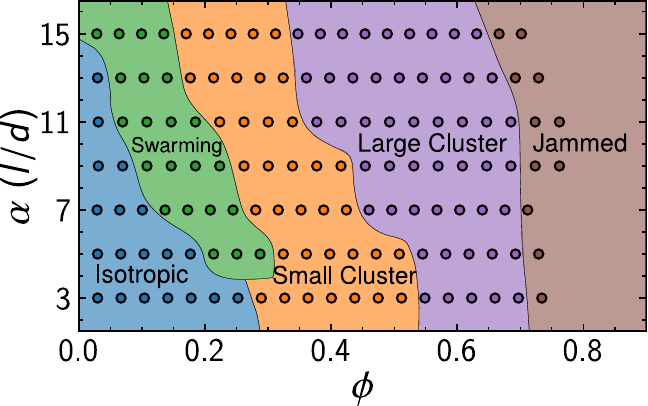}
    \caption{\textbf{Simulation state diagram of self-propelled rods as a function of rod $\alpha$ and $\phi$.} For each $\alpha$, we perform simulations at 20 periodically increasing area fractions. After which, distinct states of collective dynamics are identified based on the cluster size distribution (CSD). The diagram features isotropic, swarming, small, and large clusters, and jammed states as the $\phi$ increases. The diagram demonstrates how shape anisotropy and density together govern the emergence of collective behavior in active rod systems. The turbulent state is not observed in simulations due to the absence of hydrodynamic interactions. }
    \label{fig:simstatediag}
\end{figure}

\begin{figure}[!hbt]
    \centering
    \includegraphics[width=0.75\linewidth]{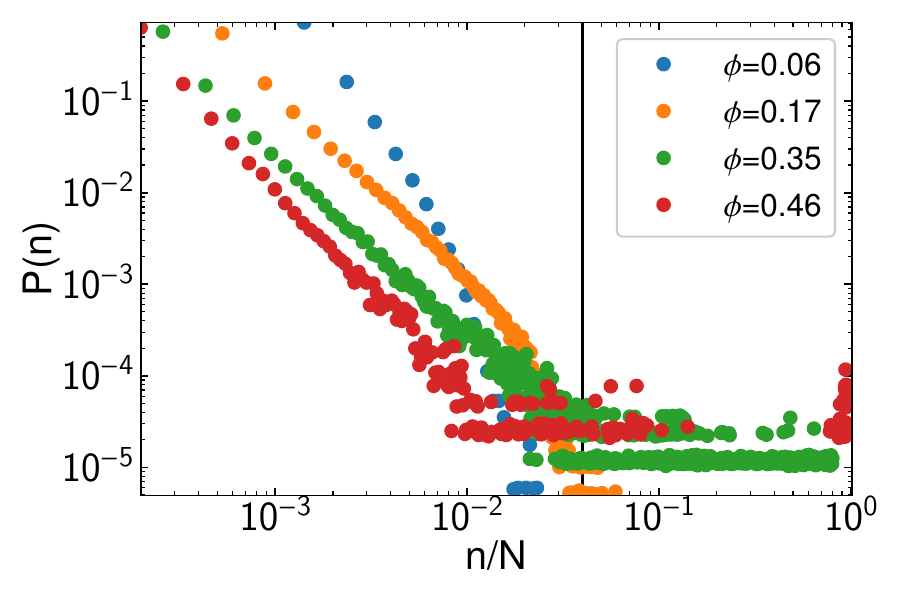}
    \caption{\textbf{Cluster size distribution $P(n)$ for simulations of rods with  $\alpha = 7$ at increasing $\phi$.} The x-axis is normalized by the total number of rods $N$. At low density ($\phi = 0.06$), the distribution is steeply decaying, consistent with an isotropic state. As $\phi$ increases, cluster sizes grow, and the distribution widens. At $\phi = 0.46$, a bimodal distribution emerges, marking the onset of the large-cluster state. The vertical dashed line at $n/N = 0.04$ is the threshold that distinguishes the isotropic state from the swarming state.}
    \label{fig:csdlow}
\end{figure}

\begin{figure}[!hbt]
    \centering
    \includegraphics[width=0.75\linewidth]{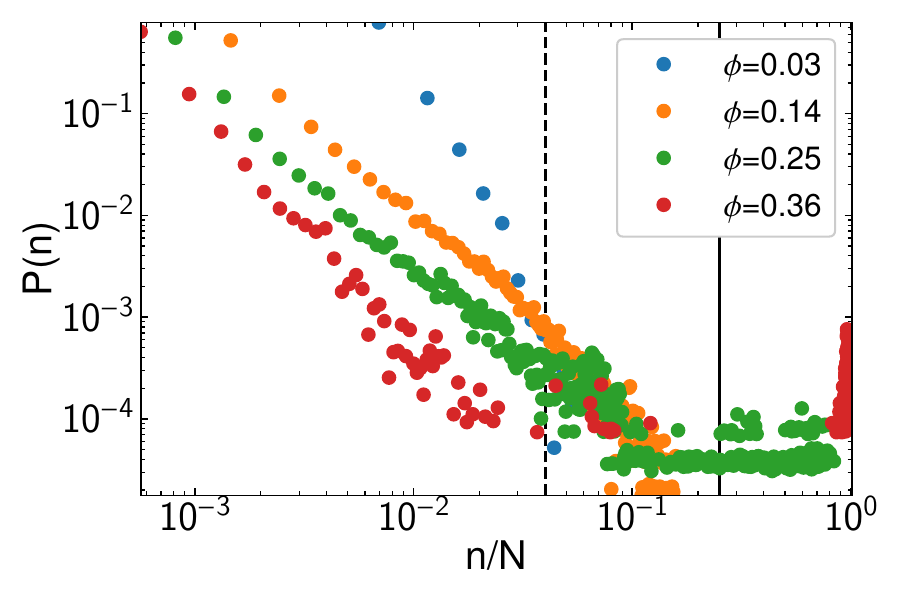}
    \caption{\textbf{Cluster size distribution $P(n)$ for simulations of rods with $\alpha = 15$ at increasing $\phi$.} The x-axis is normalized by the total number of rods $N$. Even at low density ($\phi = 0.03$), the maximum cluster size has crossed the 4\% threshold, indicating the swarming state. As $\phi$ increases, cluster sizes grow, and the distribution widens. At as low as $\phi = 0.36$, a bimodal distribution emerges, indicating the presence of system-spanning large clusters. The vertical dashed line at $n/N = 0.04$ is the threshold used to distinguish the isotropic state from the swarming state. Whereas the solid black line at $n/N = 0.25$ is the threshold used to separate the swarming and small cluster states.}
    \label{fig:csd15}
\end{figure}

\begin{figure}[!hbt]
    \centering
    \includegraphics[width=0.75\linewidth]{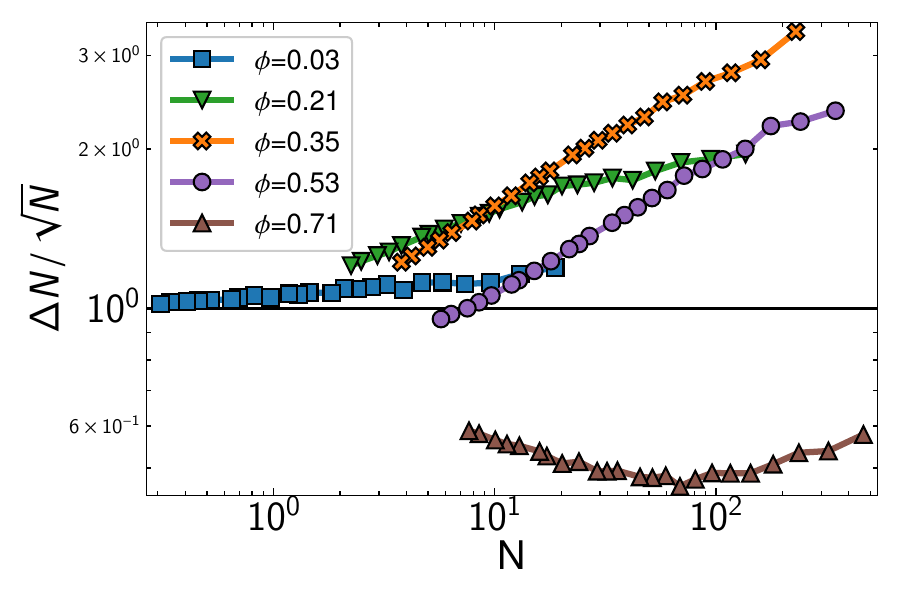}
    \caption{\textbf{Normalized number fluctuations ($\Delta N/\sqrt{\langle N \rangle}$) plotted against the mean number of rods ($\langle N \rangle$) in each subsystem from simution results for rods with $\alpha = 7$.} The horizontal solid black line at $\Delta N/\sqrt{\langle N \rangle} = 1$ represents the baseline for an isotropic equilibrium system. At low area fractions $\phi=0.03$, the fluctuations remain close to this baseline, whereas, at $\phi =0.35$, the fluctuations increase to a maximum, indicating the emergence of giant number fluctuations associated with collective motion.}
    \label{fig:numfluctsim}
\end{figure}

\begin{figure}[!hbt]
    \centering
    \includegraphics[width=0.7\linewidth]{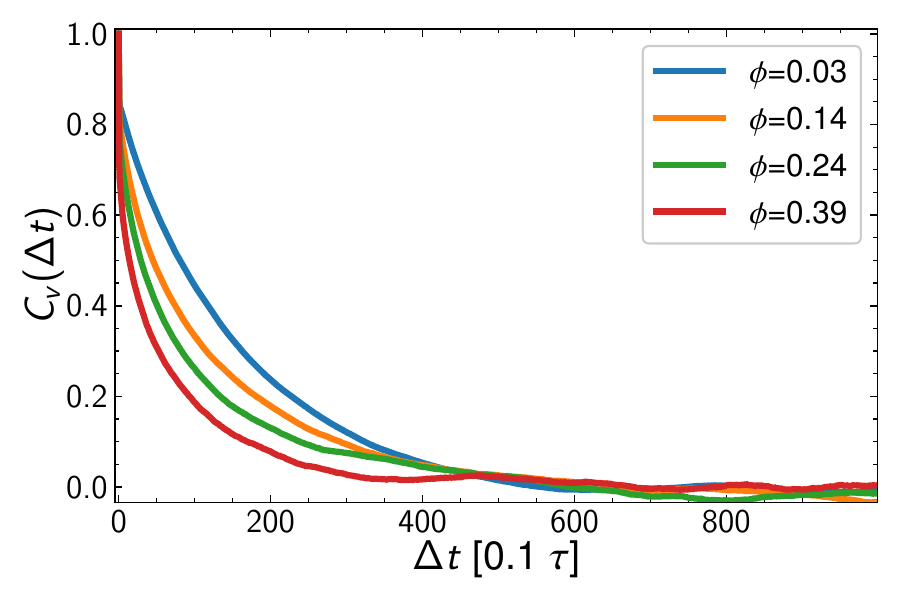}
    \caption{\textbf{Velocity autocorrelation function (VACF), $C_v(\Delta t)$, obtained from simulations of self-propelled rods with $\alpha = 7$ at various area fractions.} The VACF is normalized such that $C_v(0) = 1$. As the $\phi$ increases, steric interactions between rods become more frequent, leading to a more rapid decay of velocity correlations.}

    \label{fig:vacfsim}
\end{figure}

\begin{figure}[!hbt]
    \centering
    \includegraphics[width=\linewidth]{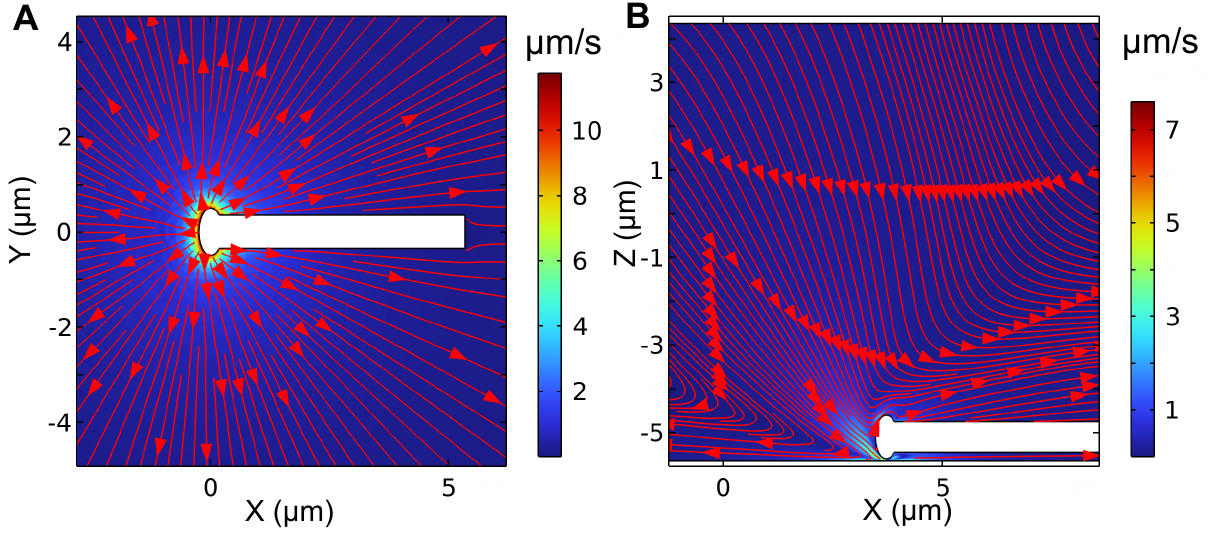}
    \caption{{\textbf{COMSOL simulations.}} (A) Flow fields obtained from COMSOL simulations ($\mathbf{u}_{\text{total}}$) along the XY plane through the center of the rod including the phoretic effects on tracers. (B) Osmotic flows ($\mathbf{u}$) along the XZ plane demonstrating the effect of the wall.}
    \label{fig:comsol}
\end{figure}
}
}

\clearpage
\newpage

\begin{table}[h!]
\centering
\caption{Zeta potentials of tracer particles used in flow field experiments.}
\begin{tabular}{|l|c|c|}
\hline
\textbf{Tracer type} &
\textbf{Diameter ($\mu$m)} &
\textbf{Zeta potential (mV) in fuel solution} \\
\hline
\textcolor{black}{Bare silica spheres (SiO\textsubscript{2})} & \textcolor{black}{0.54} & \textcolor{black}{-63} \\
\textcolor{black}{Bare silica spheres (SiO\textsubscript{2})} & \textcolor{black}{0.70} & \textcolor{black}{-62.6} \\
\textcolor{black}{Bare silica spheres (SiO\textsubscript{2})} & \textcolor{black}{0.95} & \textcolor{black}{-62.2} \\
\textcolor{black}{Coated silica spheres (SiO\textsubscript{2})} & \textcolor{black}{1.04} & \textcolor{black}{-23.9} \\
\textcolor{black}{TPM spheres} & \textcolor{black}{0.75} & \textcolor{black}{-60.8} \\
\hline
\end{tabular}
\label{tab:zeta}
\end{table}

\begin{table}[H]
\centering
\caption{Parameters used in Brownian dynamics simulations of self-propelled rods.}
\begin{tabular}{|p{3cm}|c|p{4cm}|p{6cm}|}
\hline
\textbf{Parameter} & \textbf{Symbol} & \textbf{Value} & \textbf{Description} \\
\hline
Bead diameter & $\sigma$ & 1.0 & Simulation length scale \\
Bond spring constant & $k_{\text{bond}}$ & 
$1000\,k_BT/\sigma^2$ (for $\alpha \leq 10$), $2000\,k_BT/\sigma^2$ (for $\alpha > 10$) & Ensures fixed spacing between adjacent beads; stiffer for longer rods \\
Angle spring constant & $k_{\text{angle}}$ & Same as $k_{\text{bond}}$ & Prevents bending; keeps rods straight \\
Propulsion force & $F_p$ & $10\,k_BT$ & Applied to head bead along its orientation vector \\
Simulation box size & $L$ & $300\,\sigma \times 300\,\sigma$ & Simulation domain with periodic boundaries \\
Time step & $\Delta t$ & $10^{-5}\,\tau$ & Integration step for overdamped Langevin equation \\
Temperature & $T$ & 0.1 & System temperature in LJ units \\
Translational drag & $\gamma_t$ & Set based on $\alpha$ & Set to reproduce correct anisotropic diffusion \cite{yangBeadRodComparison2017} \\
Rotational drag & $\gamma_r$ & Set based on $\alpha$ & Matches rotational diffusion theory \cite{tiradoComparisonTheoriesTranslational1984} \\
Repulsion strength & $\epsilon$ & $100\,k_BT$ & WCA interaction strength between rods \\
\hline
\end{tabular}
\label{tab:sim_params}
\end{table}

\begin{table}[h!]
\centering
\caption{COMSOL model parameters.}
\label{tab:params}
\begin{tabular}{|l|c|c|}
\hline
\textbf{Parameter} & \textbf{Symbol} & \textbf{Value} \\
\hline
Solute diffusivity               & \(D\)              & \(10^{-9}~\mathrm{m^2\,s^{-1}}\) \\
Fluid viscosity                  & \(\mu\)           & \(10^{-3}~\mathrm{Pa\,s}\) \\
Active surface flux              & \(\kappa\)         & \(10^{-6}~\mathrm{mol\,m^{-2}\,s^{-1}}\) \\
Catalytic head mobility          & \(b_c\)            & \(10^{-8}~\mathrm{m^5\,s^{-1}\,mol^{-1}}\) \\
Inert rod mobility               & \(b_i\)            & \(0.5\times10^{-8}~\mathrm{m^5\,s^{-1}\,mol^{-1}}\) \\
Wall mobility                    & \(b_{\mathrm{w}}\) & \(10^{-8}~\mathrm{m^5\,s^{-1}\,mol^{-1}}\) \\
Tracer mobility               & \(b_{tracer}\)            & \(-10^{-8}~\mathrm{m^5\,s^{-1}\,mol^{-1}}\) \\
\hline
\end{tabular}
\end{table}

\newpage

\clearpage
\section{Captions for the Supplementary Movies}
\begin{itemize}

\item \textbf{Movie S1:} \textbf{{Active Brownian motion of light-driven rods at low concentration.}} Time-lapse movie of combined fluorescence and bright-field microscopy showing the isotropic phase at low area fractions ($\phi \leq 0.08$) for rods with different aspect ratios ($\alpha = 4.5$, $\alpha = 7.5$, and $\alpha = 15.1$). The overlaid trajectories illustrate the characteristic \YS{active Brownian motion} of rods.

\YS{\item \textbf{Movie S2:} \textbf{{Motion of uncoated and polymer-coated silica tracers around an immobilized active rod.}} Combined bright-field and fluorescence microscopy movie showing tracer trajectories driven by an immobilized active rod (\(\alpha = 7.5\))  on a glass substrate for uncoated and coated silica tracers. \textbf{Left:} Overlaid trajectories show that bare (uncoated) silica tracers (0.95~$\mu$m) experience strong repulsion in front of the catalytic head and accumulation near the rear. The movie shows illumination with the green light turned off for 1~s and on for 3~s. The recording was captured at 100~fps and played back at 20~fps. \textbf{Right:} For pNIPAM-coated silica tracers (1.04~$\mu$m), the trajectories reveal isotropic repulsion around the rod head. The movie shows continuous illumination with the green light on for 20~s, captured and played back at 100~fps.}

\item \textbf{Movie S3:} \textbf{{Dynamic evolution of collective polar order and clustering in the swarming state.}} Combined fluorescence and bright-field microscopy movie of $\alpha = 7.5$ at $\phi = 0.11$ showing the evolution of rod clusters over time, with overlaid trajectories. The plot on the right displays the corresponding local polar order parameter of five tracked rods as the cluster forms, moves collectively, and eventually separates.

\item \textbf{Movie S4:} \textbf{{Swarming phase.}} Combined fluorescence and bright-field microscopy movie of rods during the swarming phase at area fractions $\phi = 0.23$ and $\phi = 0.16$ for aspect ratios $\alpha = 7.5$ (left) and $\alpha = 15.1$ (right). Left panel: Rods with aspect ratio $\alpha = 7.5$, imaged at 100× magnification. Overlaid trajectories in yellow and magenta show two distinct swarms undergoing cluster formation, merging, and splitting over time. Right panel: Rods with $\alpha = 15.1$, imaged with a large field of view (40× magnification), show a swarming state. A higher-resolution video acquired at 100× magnification, with overlaid trajectories of two swarms in red and cyan, highlights the dynamics of swarm formation and motion.

\item \textbf{Movie S5:} \textbf{{Turbulent phase.}} Combined fluorescence and bright-field movie showing the evolution of the turbulent phase for rods with $\alpha = 7.5$ at $\phi = 0.39$. A magnified view (right panel), corresponding to the region marked by a white dotted line in the left panel, shows vortex formation with overlaid velocity vectors. The vortex center is highlighted with a red star.

\item \textbf{Movie S6:} \textbf{{Large cluster phase.}} Combined fluorescence and bright-field movie showing phase separation for $\alpha = 4.5$ at $\phi \approx 0.6$, and the large cluster phase for $\alpha = 7.5$ and $\alpha = 15.1$ at $\phi = 0.72$ and $\phi = 0.65$, respectively.

\item \textbf{Movie S7:} \textbf{{Jammed phase.}}  Combined fluorescence and bright-field movie showing the jammed phase at $\phi \approx 0.9$ for rods with  $\alpha = 7.5$.

\item \textbf{Movie S8:} \textbf{{Transient cluster phase.}}  Combined fluorescence and bright-field movie showing the transient cluster phase at $\phi = 0.27$ for rods with  $\alpha = 4.5$. The transient nature of the clusters is demonstrated by the overlaid color that shows temporal variations in cluster size ($n \geq 10$).

\item \textbf{Movie S9:} \textbf{{Flocking phase.}} Combined fluorescence and bright-field movie showing the flocking phase for longer rods ($\alpha = 15.1$) at $\phi = 0.48$. A magnified view (right panel), corresponding to the region marked by a white dotted line in the left panel, highlights the merging of smaller swarms with different orientations into larger, coherently moving flocks, indicated by red arrows.

\item \textbf{Movie S10:} \textbf{{Simulation of active rods collective states.}} Simulation video showing various states of collective dynamics of active rods at different $\phi$ and $\alpha$.

\end{itemize}

\clearpage

\begin{thebibliography}{10}

\bibitem{bechingerActiveParticlesComplex2016}
C.~Bechinger, R.~Di~Leonardo, H.~L{\"o}wen, C.~Reichhardt, G.~Volpe, G.~Volpe, Active {{Particles}} in {{Complex}} and {{Crowded Environments}}.
\newblock {\it Rev. Mod. Phys.\/} {\bf 88},  045006 (2016).

\bibitem{Gompper_2025}
G.~Gompper, {\it et~al.\/}, The 2025 motile active matter roadmap.
\newblock {\it J. Phys. Condens. Matter\/} {\bf 37},  143501 (2025).

\bibitem{yan2016reconfiguring}
J.~Yan, M.~Han, J.~Zhang, C.~Xu, E.~Luijten, S.~Granick, Reconfiguring active particles by electrostatic imbalance.
\newblock {\it Nat. Mater.\/} {\bf 15},  1095 (2016).

\bibitem{theurkauff2012dynamic}
I.~Theurkauff, C.~Cottin-Bizonne, J.~Palacci, C.~Ybert, L.~Bocquet, Dynamic clustering in active colloidal suspensions with chemical signaling.
\newblock {\it Phys. Rev. Lett.\/} {\bf 108},  268303 (2012).

\bibitem{vutukuri2020active}
H.~R. Vutukuri, {\it et~al.\/}, Active particles induce large shape deformations in giant lipid vesicles.
\newblock {\it Nature\/} {\bf 586}, ~52 (2020).

\bibitem{palacciLivingCrystalsLightActivated2013}
J.~Palacci, S.~Sacanna, A.~P. Steinberg, D.~J. Pine, P.~M. Chaikin, Living {{Crystals}} of {{Light-Activated Colloidal Surfers}}.
\newblock {\it Science\/} {\bf 339},  936 (2013).

\bibitem{catesMotilityInducedPhaseSeparation2015}
M.~E. Cates, J.~Tailleur, Motility-{{Induced Phase Separation}}.
\newblock {\it Annu. Rev. Condens. Matter Phys.\/} {\bf 6},  219 (2015).

\bibitem{weitzSelfpropelledRodsExhibit2015}
S.~Weitz, A.~Deutsch, F.~Peruani, Self-propelled rods exhibit a phase-separated state characterized by the presence of active stresses and the ejection of polar clusters.
\newblock {\it Phys. Rev. E\/} {\bf 92},  012322 (2015).

\bibitem{lushiFluidFlows2014}
E.~Lushi, H.~Wioland, R.~E. Goldstein, Fluid flows created by swimming bacteria drive self-organization in confined suspensions.
\newblock {\it Proc. Nat. Acad. Sci. USA\/} {\bf 111},  9733 (2014).

\bibitem{colinChemotacticBehaviourEscherichia2019}
R.~Colin, K.~Drescher, V.~Sourjik, Chemotactic behaviour of {{Escherichia}} coli at high cell density.
\newblock {\it Nat. Commun.\/} {\bf 10},  5329 (2019).

\bibitem{peruaniCollectiveMotionNonequilibrium2012}
F.~Peruani, J.~Starru{\ss}, V.~Jakovljevic, L.~{S{\o}gaard-Andersen}, A.~Deutsch, M.~B{\"a}r, Collective {{Motion}} and {{Nonequilibrium Cluster Formation}} in {{Colonies}} of {{Gliding Bacteria}}.
\newblock {\it Phys. Rev. Lett.\/} {\bf 108},  098102 (2012).

\bibitem{zhangCollectiveMotionDensity2010}
H.~P. Zhang, A.~Be’er, E.~L. Florin, H.~L. Swinney, Collective motion and density fluctuations in bacterial colonies.
\newblock {\it Proc. Nat. Acad. Sci. USA\/} {\bf 107},  13626 (2010).

\bibitem{zantopEmergentCollectiveDynamics2022}
A.~W. Zantop, H.~Stark, Emergent collective dynamics of pusher and puller squirmer rods: Swarming, clustering, and turbulence.
\newblock {\it Soft Matter\/} {\bf 18},  6179 (2022).

\bibitem{peruaniNonequilibriumClusteringSelfpropelled2006}
F.~Peruani, A.~Deutsch, M.~Bär, Nonequilibrium clustering of self-propelled rods.
\newblock {\it Phys. Rev. E\/} {\bf 74},  030904 (2006).

\bibitem{grossmann2020particle}
R.~Gro{\ss}mann, I.~S. Aranson, F.~Peruani, A particle-field approach bridges phase separation and collective motion in active matter.
\newblock {\it Nat. Commun.\/} {\bf 11},  5365 (2020).

\bibitem{broker2024collective}
S.~Br{\"o}ker, M.~te~Vrugt, R.~Wittkowski, Collective dynamics and pair-distribution function of active brownian ellipsoids in two spatial dimensions.
\newblock {\it Commun. Phys.\/} {\bf 7},  238 (2024).

\bibitem{ginelliLargeScaleCollectiveProperties2010}
F.~Ginelli, F.~Peruani, M.~Bär, H.~Chaté, Large-{{Scale Collective Properties}} of {{Self-Propelled Rods}}.
\newblock {\it Phys. Rev. Lett.\/} {\bf 104},  184502 (2010).

\bibitem{baskaran2008hydrodynamics}
A.~Baskaran, M.~C. Marchetti, Hydrodynamics of self-propelled hard rods.
\newblock {\it Phys. Rev. E\/} {\bf 77},  011920 (2008).

\bibitem{alertActiveTurbulence2022}
R.~Alert, J.~Casademunt, J.~F. Joanny, Active {{Turbulence}}.
\newblock {\it Annu. Rev. Condens. Matter Phys.\/} {\bf 13},  143 (2022).

\bibitem{wensinkMesoscaleTurbulenceLiving2012}
H.~H. Wensink, {\it et~al.\/}, Meso-scale turbulence in living fluids.
\newblock {\it Proc. Nat. Acad. Sci. USA\/} {\bf 109},  14308 (2012).

\bibitem{narayanLongLivedGiantNumber2007}
V.~Narayan, S.~Ramaswamy, N.~Menon, Long-{{Lived Giant Number Fluctuations}} in a {{Swarming Granular Nematic}}.
\newblock {\it Science\/} {\bf 317},  105 (2007).

\bibitem{kudrolli2008swarming}
A.~Kudrolli, G.~Lumay, D.~Volfson, L.~S. Tsimring, Swarming and swirling in self-propelled polar granular rods.
\newblock {\it Phys. Rev. Lett.\/} {\bf 100},  058001 (2008).

\bibitem{vutukuri2020light}
H.~R. Vutukuri, M.~Lisicki, E.~Lauga, J.~Vermant, Light-switchable propulsion of active particles with reversible interactions.
\newblock {\it Nat. Commun.\/} {\bf 11},  2628 (2020).

\bibitem{barSelfPropelledRodsInsights2020}
M.~B{\"a}r, R.~Gro{\ss}mann, S.~Heidenreich, F.~Peruani, Self-{{Propelled Rods}}: {{Insights}} and {{Perspectives}} for {{Active Matter}}.
\newblock {\it Annu. Rev. Condens. Matter Phys.\/} {\bf 11},  441 (2020).

\bibitem{yangSwarmBehaviorSelfpropelled2010}
Y.~Yang, V.~Marceau, G.~Gompper, Swarm behavior of self-propelled rods and swimming flagella.
\newblock {\it Phys. Rev. E\/} {\bf 82},  031904 (2010).

\bibitem{vutukuriDynamicSelforganizationSidepropelling2016}
H.~R. Vutukuri, Z.~Preisler, T.~H. Besseling, A.~Van~Blaaderen, M.~Dijkstra, W.~T.~S. Huck, Dynamic self-organization of side-propelling colloidal rods: Experiments and simulations.
\newblock {\it Soft Matter\/} {\bf 12},  9657 (2016).

\bibitem{paxtonCatalyticNanomotorsAutonomous2004}
W.~F. Paxton, {\it et~al.\/}, Catalytic {{Nanomotors}}:\, {{Autonomous Movement}} of {{Striped Nanorods}}.
\newblock {\it J. Am. Chem. Soc.\/} {\bf 126},  13424 (2004).

\bibitem{mu2022light}
Y.~Mu, W.~Duan, K.~Y. Hsu, Z.~Wang, W.~Xu, Y.~Wang, Light-activated colloidal micromotors with synthetically tunable shapes and shape-directed propulsion.
\newblock {\it ACS Appl. Mater. Interfaces\/} {\bf 14},  57113 (2022).

\bibitem{kuijk2011synthesis}
A.~Kuijk, A.~Van~Blaaderen, A.~Imhof, Synthesis of monodisperse, rodlike silica colloids with tunable aspect ratio.
\newblock {\it J. Am. Chem. Soc.\/} {\bf 133},  2346 (2011).

\bibitem{thielicke2021particle}
W.~Thielicke, R.~Sonntag, Particle image velocimetry for matlab: Accuracy and enhanced algorithms in pivlab.
\newblock {\it J. Open Res. Softw.\/} {\bf 9}, ~12 (2021).

\bibitem{marchetti2013hydrodynamics}
M.~C. Marchetti, {\it et~al.\/}, Hydrodynamics of soft active matter.
\newblock {\it Rev. Mod. Phys.\/} {\bf 85},  1143 (2013).

\bibitem{andersonColloidTransportInterfacial1989}
J.~L. Anderson, Colloid {{Transport}} by {{Interfacial Forces}}.
\newblock {\it Annu. Rev. Fluid Mech.\/} {\bf 21}, ~61 (1989).

\bibitem{spagnolieHydrodynamicsSelfpropulsionBoundary2012}
S.~E. Spagnolie, E.~Lauga, Hydrodynamics of self-propulsion near a boundary: Predictions and accuracy of far-field approximations.
\newblock {\it J. Fluid Mech.\/} {\bf 700},  105 (2012).

\bibitem{drescherFluidDynamicsNoise2011}
K.~Drescher, J.~Dunkel, L.~H. Cisneros, S.~Ganguly, R.~E. Goldstein, Fluid dynamics and noise in bacterial cell--cell and cell--surface scattering.
\newblock {\it Proc. Nat. Acad. Sci. USA\/} {\bf 108},  10940 (2011).

\bibitem{carrasco2025characterization}
C.~Carrasco, Q.~Martinet, Z.~Shen, J.~Lintuvuori, J.~Palacci, A.~Aubret, Characterization of nonequilibrium interactions of catalytic microswimmers using phoretically responsive nanotracers.
\newblock {\it ACS Nano\/} {\bf 19},  11133 (2025).

\bibitem{liuDensityFluctuationsEnergy2021}
Z.~Liu, W.~Zeng, X.~Ma, X.~Cheng, Density fluctuations and energy spectra of {{3D}} bacterial suspensions.
\newblock {\it Soft Matter\/} {\bf 17},  10806 (2021).

\bibitem{qiEmergenceActiveTurbulence2022}
K.~Qi, E.~Westphal, G.~Gompper, R.~G. Winkler, Emergence of active turbulence in microswimmer suspensions due to active hydrodynamic stress and volume exclusion.
\newblock {\it Commun. Phys.\/} {\bf 5}, ~1 (2022).

\bibitem{pengImagingEmergenceBacterial2021}
Y.~Peng, Z.~Liu, X.~Cheng, Imaging the emergence of bacterial turbulence: {{Phase}} diagram and transition kinetics.
\newblock {\it Sci. Adv.\/} {\bf 7},  eabd1240 (2021).

\bibitem{beerPhaseDiagramBacterial2020}
A.~Be’er, {\it et~al.\/}, A phase diagram for bacterial swarming.
\newblock {\it Commun. Phys.\/} {\bf 3}, ~66 (2020).

\bibitem{nishiguchiLongrangeNematicOrder2017}
D.~Nishiguchi, K.~H. Nagai, H.~Chat{\'e}, M.~Sano, Long-range nematic order and anomalous fluctuations in suspensions of swimming filamentous bacteria.
\newblock {\it Phys. Rev. E\/} {\bf 95},  020601 (2017).

\bibitem{4tudata}
Y.~Shelke, A. Nair S, H.~R. Vutukuri, Data and analysis codes, 4TU.ResearchData (2026); https://doi.org/10.4121/67c7ba9a-29a9-40f4-a260-5a4ba2d35b77

\bibitem{zhaoOnestepSynthesisMicronsized2025}
Z.~Zhao, {\it et~al.\/}, One-step synthesis of micron-sized silica particles by continuous dropwise addition: {{The}} effect of reaction parameters on particle size.
\newblock {\it Mater. Chem. Phys.\/} {\bf 335},  130525 (2025).

\bibitem{ershovTrackMate7Integrating2022}
D.~Ershov, {\it et~al.\/}, {{TrackMate}} 7: Integrating state-of-the-art segmentation algorithms into tracking pipelines.
\newblock {\it Nat. Methods\/} {\bf 19},  829 (2022).

\bibitem{scikit-learn}
F.~Pedregosa, {\it et~al.\/}, Scikit-learn: Machine learning in {P}ython.
\newblock {\it J. Mach. Learn. Res.\/} {\bf 12},  2825 (2011).

\bibitem{LAMMPS}
A.~P. Thompson, {\it et~al.\/}, {LAMMPS} - a flexible simulation tool for particle-based materials modeling at the atomic, meso, and continuum scales.
\newblock {\it Comp. Phys. Comm.\/} {\bf 271},  108171 (2022).

\bibitem{tiradoComparisonTheoriesTranslational1984}
M.~M. Tirado, C.~L. Martínez, J.~G. de~la Torre, Comparison of theories for the translational and rotational diffusion coefficients of rod‐like macromolecules. {{Application}} to short {{DNA}} fragments.
\newblock {\it J. Chem. Phys.\/} {\bf 81},  2047 (1984).

\bibitem{michaud2011mdanalysis}
N.~Michaud-Agrawal, E.~J. Denning, T.~B. Woolf, O.~Beckstein, Mdanalysis: a toolkit for the analysis of molecular dynamics simulations.
\newblock {\it J. Comput. Chem.\/} {\bf 32},  2319 (2011).

\bibitem{yangBeadRodComparison2017}
K.~Yang, C.~Lu, X.~Zhao, R.~Kawamura, From bead to rod: {{Comparison}} of theories by measuring translational drag coefficients of micron-sized magnetic bead-chains in {{Stokes}} flow.
\newblock {\it PLOS ONE\/} {\bf 12},  e0188015 (2017).

\bibitem{bitterInterfacialConfinedColloidal2017}
J.~L. Bitter, Y.~Yang, G.~Duncan, H.~Fairbrother, M.~A. Bevan, Interfacial and {{Confined Colloidal Rod Diffusion}}.
\newblock {\it Langmuir\/} {\bf 33},  9034 (2017).

\bibitem{sokolovConcentrationDependenceCollective2007a}
A.~Sokolov, I.~S. Aranson, J.~O. Kessler, R.~E. Goldstein, Concentration {{Dependence}} of the {{Collective Dynamics}} of {{Swimming Bacteria}}.
\newblock {\it Phys. Rev. Lett.\/} {\bf 98},  158102 (2007).

\bibitem{hardikarOsmoticPhoreticCompetition2024}
A.~V. Hardikar, A.~W. Hauser, T.~M. Hopkins, S.~Sacanna, P.~M. Chaikin, Osmotic and phoretic competition explains chemotaxic assembly and sorting.
\newblock {\it Proc. Nat. Acad. Sci. USA\/} {\bf 121},  e2410840121 (2024).

\bibitem{katuriInferringNonequilibriumInteractions2021}
J.~Katuri, W.~E. Uspal, M.~N. Popescu, S.~S{\'a}nchez, Inferring non-equilibrium interactions from tracer response near confined active {{Janus}} particles.
\newblock {\it Sci. Adv.\/} {\bf 7},  eabd0719 (2021).


\end{thebibliography}

\newbox\hidenotes



\end{document}